\documentclass[aps,prl,preprint,superscriptaddress]{revtex4-1}

\usepackage{graphicx}
\usepackage{amsmath}
\usepackage{amssymb}
\usepackage{slashed}

\begin{document}


\title{On finite volume effects \\ in the chiral extrapolation of baryon masses}


\author{M.F.M. Lutz}
\affiliation{GSI Helmholtzzentrum f\"ur Schwerionenforschung GmbH, \\Planckstra\ss e 1, 64291 Darmstadt, Germany}
\author{R. Bavontaweepanya}
\affiliation{Mahidol University, Bangkok 10400,  Thailand}
\author{C. Kobdaj}
\affiliation{Suranaree University of Technology, Nakhon Ratchasima, 30000, Thailand}
\author{K. Schwarz}
\affiliation{GSI Helmholtzzentrum f\"ur Schwerionenforschung GmbH, \\Planckstra\ss e 1, 64291 Darmstadt, Germany}
\date{\today}

\begin{abstract}
We perform an analysis of the QCD lattice data on the baryon octet and decuplet masses based on the relativistic
chiral Lagrangian. The baryon self energies are computed in a finite volume at next-to-next-to-next-to leading
order (N$^3$LO), where the dependence on the physical meson and baryon masses is kept.
The number of free parameters
is reduced significantly down to 12 by relying on large-$N_c$ sum rules.
Altogether we describe accurately more than 220 data points from six different lattice groups, BMW, PACS-CS, HSC, LHPC,
QCDSF-UKQCD and NPLQCD. Values for all counter terms relevant at N$^3$LO are predicted.
In partic\-ular we extract a pion-nucleon sigma term of 39$_{-1}^{+2}$  MeV and a strangeness sigma term of the nucleon of
$\sigma_{sN} = 84^{+ 28}_{-\;4}$ MeV. The flavour SU(3) chiral limit of the baryon octet and decuplet masses is determined
with $( 802 \pm 4 )$ MeV and $(1103 \pm 6)$ MeV. Detailed predictions for the baryon masses as currently evaluated
by the ETM lattice QCD group are made.
\end{abstract}


\maketitle

\section{Introduction}
\label{sec:intro}
Various unquenched three-flavour simulations  for the pion- and kaon-mass dependence of the
baryon ground state masses are available \cite{LHPC2008,PACS-CS2008,HSC2008,BMW2008,Alexandrou:2009qu,Durr:2011mp,Bietenholz:2011qq,WalkerLoud:2011ab}.
Such data are expected to determine the low-energy constants of the three-flavour chiral Lagrangian
formulated with the baryon octet and decuplet fields. A precise knowledge of the latter parameters is crucial
for a profound understanding of meson-baryon scattering data based on coupled-channel dynamics as derived from
the chiral Lagrangian (see e.g. \cite{Kaiser:1995eg,Lutz:2001yb,Ikeda:2012au,Bruns:2010sv}).

The chiral extrapolation of baryon masses with strangeness content is
discussed extensively in the literature \cite{Jenkins1991,WalkerLoud:2004hf,Tiburzi:2004rh,Semke2005,Frink2006,Semke2007,Tiburzi:2008bk,Jiang:2009fa,MartinCamalich:2010fp,PhysRevD.81.014503,Geng:2011wq,MartinCamalich:2010zz,Semke:2011ez,Semke:2012gs,Lutz:2012mq,Bruns:2012eh,Ren:2013dzt,Ren:2013oaa}.
The convergence properties of a strict chiral expansion for the baryon masses with three light flavours are very poor,
if existing at all for physical strange quark masses.
Different strategies how to perform partial summations or phenomenological adaptations are investigated by several
groups \cite{Semke2005,Frink2006,Semke2007,MartinCamalich:2010fp,PhysRevD.81.014503}.
A straightforward application of chiral perturbation theory to QCD lattice simulations appears
futile (see e.g. \cite{LHPC2008,PACS-CS2008,Alexandrou:2009qu,WalkerLoud:2011ab,MartinCamalich:2010fp}).

Recently, systematic analyses of lattice data at N$^3$LO were performed \cite{Semke:2011ez,Semke:2012gs,Lutz:2012mq,Bruns:2012eh,Ren:2013dzt,Ren:2013oaa},
where different versions and renormalization schemes for the relativistic chiral Lagrangian were applied. While
in \cite{Semke:2011ez,Semke:2012gs,Lutz:2012mq} the $\chi \overline{MS}$ scheme of \cite{Semke2005} was used,
the EOMS of \cite{Gegelia:1999gf} was used in \cite{Ren:2013dzt,Ren:2013oaa}. Both schemes protect the analytic
structure of the one-loop contributions in the baryon self energies as requested by micro causality. In contrast,
the IR scheme of \cite{Becher:1999he} was applied in \cite{Bruns:2012eh}. All three schemes are consistent with the
heavy-baryon expansion, and therefore, constitute partic\-ular summations of higher order terms. No ad-hoc form factor
or cutoff is used in any of those works.

While the works \cite{Bruns:2012eh,Ren:2012aj,Ren:2013dzt,Ren:2013oaa} consider either the baryon octet or the baryon
decuplet masses,  a simultaneous description of all baryon states is achieved
in \cite{Semke:2011ez,Semke:2012gs,Lutz:2012mq}. There is a further difference as to how the octet
and decuplet fields are coupled to the  Goldstone  bosons. In \cite{Semke:2011ez,Semke:2012gs,Lutz:2012mq}
the minimal and traditional form that was shown to be compatible with empirical decay properties of the baryon
decuplet decays (see e.g. \cite{Semke2005}), was used. The so called 'consistent' coupling form suggested
in \cite{Pascalutsa:1998pw,Pascalutsa:1999zz} was applied in \cite{Ren:2013dzt,Ren:2013oaa}. The latter form was not
used in \cite{Semke:2011ez,Semke:2012gs,Lutz:2012mq} since it does not seem to be compatible with the
empirical decay pattern of the decuplet states.

The available lattice data were reproduced at different levels. A first simultaneous description of
the BMW, PACS-CS, HSC, LHPC and QCDSF-UKQCD data was achieved in \cite{Semke:2011ez,Semke:2012gs,Lutz:2012mq}.
The 6 parameter scenario of \cite{Semke:2011ez,Semke:2012gs,Lutz:2012mq} demonstrated that it was possible to
predict for instance the HSC and QCDSF-UKQCD data in terms of the BMW and PACS-CS data. The idea of a simultaneous
description of the lattice data was taken up in \cite{Ren:2013dzt}, where however only the baryon octet data
of various lattice groups were fitted in terms of the 19 parameters relevant at N$^3$LO in the baryon octet sector.
As compared to \cite{Semke:2011ez,Semke:2012gs,Lutz:2012mq} the number of fit parameters increased significantly. This
is in part, since in the 6 parameter scenario of \cite{Semke:2011ez,Semke:2012gs,Lutz:2012mq} the symmetry conserving
counter terms that enter at N$^3$LO were not considered. It is difficult to determine them with a subset of
lattice data that have sufficiently large lattice volumes so that finite volume corrections can be neglected. Only
when considering lattice data with smaller volumes, like the ones provided by QCDSF-UKQCD or
NPLQCD, such counter terms can possibly be determined reliably. A first study of finite volume effects at the
N$^3$LO was performed in \cite{Ren:2012aj}, however, with only partial success at smaller volumes.
In the follow-up work \cite{Ren:2013dzt} it was emphasized that in
a small lattice volume the baryon decuplet degrees of freedom are needed in a description of the baryon octet masses.
Still, even the consideration of that effect did not yet lead to a satisfactory description of the QCDSF-UKQCD results
on the 24$^3$ and 16$^3$ lattice.

Our work is based on the relativistic chiral Lagrangian with baryon octet and decuplet fields where
effects at N$^3$LO are considered systematically. Here the leading order (LO)
corresponds to the chiral limit of the baryon masses and the NLO effects are linear in the quark masses. The details
of the approach are published
in  \cite{Semke2005,Semke2007,Semke:2011ez}. A crucial element of our scheme is the use of physical masses in the
one-loop contribution to the baryon self energies. Furthermore, the low-energy constants required at N$^3$LO are estimated by sum rules that follow from QCD in the
limit of a large number of colors ($N_c$) \cite{LutzSemke2010,Semke:2011ez,WalkerLoud:2011ab}. The approach was successfully tested against the available lattice
data on the nucleon and omega masses of the BMW group \cite{BMW2008}. Adjusting eight low-energy constants to the empirical baryon masses we
fitted the remaining 6 parameters to the BMW data for the nucleon and omega. In turn we obtained results \cite{Semke:2011ez,Semke:2012gs,Lutz:2012mq}
that are in agreement with the predictions of the
PACS-CS, HSC, LHPC and QCDSF-UKQCD groups \cite{LHPC2008,PACS-CS2008,HSC2008,WalkerLoud:2011ab,Jenkins:2009wv}.

It is the aim of our present study to extend the works \cite{Semke:2011ez,Semke:2012gs,Lutz:2012mq} by considering finite
volume effects and possibly arrive at an accurate description of all QCDSF-UKQCD data on the baryon masses.
We do not consider discretization effects since in \cite{Alexandrou:2013joa,Ren:2013wxa} they were estimated to be of
minor importance only. Furthermore, we would like to test the quality of the large-N$_c$ sum rules as derived
in \cite{LutzSemke2010} for the symmetry conserving counter terms. Altogether there are 17 such counter terms contributing
to the baryon octet and decuplet masses at N$^3$LO. Owing to the sum rules of \cite{LutzSemke2010} they can be parameterized
in terms of 5 parameters only. Clearly such a study can be performed meaningfully only by a simultaneous consideration of
the baryon octet and decuplet masses.

The work is organized as follows. We first derive and present the finite volume effects appropriate for the framework
\cite{Semke:2011ez,Semke:2012gs,Lutz:2012mq}. Using large-$N_s$ sum rules we arrive at a 12 parameter scenario, which
is confronted with the lattice data on the baryon masses. As a consequence we predict a set of low-energy
constants and precise values for the baryon sigma terms. In addition detailed predictions for the baryon masses as
currently evaluated by the ETM lattice QCD group are made.


\section{Baryon self energies in a finite volume}
\label{sec:1}

We consider the chiral extrapolation of the baryon masses to unphysical quark masses. Assuming exact isospin symmetry,
the hadron masses are functions of  $m_u=m_d\equiv m$ and $m_s$. The dependence on the light quark masses may be
traded against a dependence on the pion and kaon masses. For a given lattice data set we use the lattice pion and kaon
masses to determine the quark masses as predicted by $\chi$PT at the next-to-leading
order  in a finite cubic volume with $V = L^3$. From \cite{Gasser:1984gg,Hasenfratz:1989pk} we
recall
 \begin{eqnarray}
&& m_\pi^2 =\frac{2\,B_0\,m}{f^2}\,\Big\{ f^2+  \frac{1}{2}\,\bar I_\pi -\frac{1}{6}\,\bar I_\eta
+16\,B_0\,\Big[ (2\, m+m_s)\,(2\,L_6-L_4)+  m\, (2\,L_8-L_5)\Big] \Big\}\,,
\nonumber\\
&& m_K^2 = \frac{B_0\,(m+m_s)}{f^2}\,\Big\{f^2 +
\frac{1}{3}\,\bar I_\eta
\nonumber\\
&& \qquad \qquad +\,16 \,B_0\,\Big[(2\,m+m_s)\,(2\,L_6-L_4)+\frac{1}{2}\,( m+m_s)\,(2\,L_8-L_5)\Big]
\Big\}\,,
\nonumber\\
&& m_\eta^2 = \frac{2\,B_0\,(m+2\,m_s)}{3\,f^2}\,
\Big\{ f^2 + \bar I_K- \frac{2}{3}\, \bar I_\eta
\nonumber\\
&& \qquad \qquad +\, 16\,B_0\,\Big[
(2\, m+m_s)\,(2\,L_6-L_4) +  \frac{1}{3}\, (m + 2\,m_s)\, (2\,L_8-L_5) \Big]
\Big\}
\nonumber\\
&& \quad \;\;\, + \,\frac{2\,B_0\,m}{f^2}\,\left[\frac{1}{6}\, \bar I_\eta-\frac{1}{2}\, \bar I_\pi +\frac{1}{3} \,\bar I_K\right]
+\frac{128}{9}\,\frac{B^2_0\,(m-m_s)^2}{f^2}(3\,L_7+L_8)\,,
\label{meson-masses-q4}
\end{eqnarray} 
in terms of the renormalized mesonic tadpole integrals $\bar I_Q$ with $Q =\pi, K, \eta $. The latter are
computed in the $\overline{MS}$ scheme at a renormalization scale $\mu$ with
 \begin{eqnarray}
&& \bar I_Q =\frac{m_Q^2}{(4\,\pi)^2}\,
\log \left( \frac{m_Q^2}{\mu^2}\right) + \frac{1}{4\,\pi^2}\,\sum^{\vec n \neq 0}_{\vec n \in  Z^3} \frac{m_Q}{|\vec x_n|}\,K_1(m_Q\,|\vec x_n|)\,
\qquad {\rm with} \qquad \vec x_n = L \,\vec n \,,
\label{def-tadpole}
\end{eqnarray} 

where $K_n(x)= {\rm BesselK}[n,x]$ is the modified Bessel function of second kind in the convention used in Mathematica.
For a given set of low-energy parameters $f, L_i$ and values for the pion and kaon masses, the quark masses
$B_0\,m$ and $B_0\,m_s$ together with the eta meson mass can be determined by (\ref{meson-masses-q4}) unambiguously.

There are 4 low-energy constants relevant for the meson masses: the flavour SU(3) chiral limit of the pion decay constant
$f$, together with three combinations
 \begin{eqnarray}
&& L_4 - 2\,L_6  =\phantom{-}  0.09868 \,\times \, 10^{-3}\,,
\nonumber\\
&& L_5 -2\,L_8   = -0.39208 \, \times \,10^{-3}\,,
\nonumber\\
&& L_8 + 3\,L_7  = -0.30142 \, \times \,10^{-3}\,,
\label{def-Ls}
\end{eqnarray} 
where we recall the values used in \cite{LutzSemke2010,Semke:2011ez} at the
renormalization scale $\mu = 0.77$ GeV. Note that $L_4 - 2\,L_6$ is expected to vanish in the large-$N_c$ limit.
The values are chosen such that for appropriate choices of
$B_0\,m$ and  $B_0\,m_s$ the empirical isospin averages for the pion, kaon and eta masses are recovered if
$f = 92.4$ MeV is used. According to the latest determination \cite{Bijnens:2011tb} the low-energy constants
suffer from substantial uncertainties. For instance the value of the
combination $L_5 -2\,L_8$ may be positive or negative.  Also a precise value for $f$ is not known.  Note however,
that for any given choice of $f, L_4 - 2\,L_6$ and $L_5 -2\,L_8$ the value of $L_8 + 3\,L_7$ is determined by the
eta meson mass. In the following we will initially use the value $f$ = 92.4 MeV together with (\ref{def-Ls}). The stability
of this assumption will be investigated later on.

We turn to the computation of the baryon masses. The physical mass of a baryon $M_B$ of type $B$ is determined by the condition
 \begin{eqnarray}
M_B -\Sigma_B(M_B)    = \left\{\begin{array}{ll}
\bar M_{[8]}  & \enskip {\rm for } \enskip B\in[8] = \{ N, \Lambda, \Sigma, \Xi\}\\
\bar M_{[10]}  & \enskip {\rm for }\enskip B \in[10] =\{ \Delta, \Sigma^*, \Xi^*, \Omega \}
\end{array} \right. \,,
\label{def-non-linear-system}
\end{eqnarray} 
with the self energy of the baryon $\Sigma_B(M_B)$ incorporating all tree-level and loop corrections.
Note that in the case of the decuplet a projected self energy can be used as described in \cite{Semke2005}.
The self energy is split into its tree-level and loop contribution
 \begin{eqnarray}
\Sigma_B(M_B) = \Sigma^{\rm tree-level}_B(M_B) + \Sigma^{\rm loop}_B(M_B)\,,
\label{def-tree-level}
\end{eqnarray} 
where all tree-level contributions vanish in the chiral limit. Explicit expression for the baryon self energies can
be found in the Appendix of \cite{Semke:2011ez}. There are three types of contributions considered.
The first class enters at NLO and is parameterized by the renormalized low-energy constants
$\bar b_0,\bar b_D, \bar b_F, \bar d_0$ and $\bar d_D$. The second class are wave function renormalization terms
parameterized by $\bar \zeta_0,\bar \zeta_D, \bar \zeta_F, \bar \xi_0$ and $\bar \xi_D$. Formally they turn relevant
at N$^3$LO. Finally there are the terms that are proportional to the square of the quark masses. In the octet
sector they are labelled with $\bar c_{0,.., 6}$ and in the decuplet sector they are labelled with $\bar e_{0, ..., 4}$.
All such parameters run on the ultraviolet renormalization scale $\mu$. It is determined by the request that the baryon masses
are independent on $\mu$, if expanded to second order in the quark masses. In the Appendix A the running of all
low-energy constants encountered in this work is provided.

While for the NLO parameters we do not impose
any constraints from large-$N_c$ sum rules  we do so for the parameters relevant at N$^3$LO.
A matching of the chiral interaction terms to the
large-$N_c$ operator analysis for the baryon masses in \cite{Jenkins1995} leads to the seven sum rules. We recall from \cite{Semke:2011ez}
 \begin{eqnarray}
&&\bar c_0 = \frac{1}{2} \,\bar c_1, \quad \bar c_2=-\frac{3}{2} \,\bar c_1,\quad \bar c_3=0 \,, \quad
\nonumber\\
&& \bar e_0 = 0\,, \quad \bar e_1 = -2 \,\bar c_2, \quad  \bar e_2 = 3\, \bar c_2\,, \quad \bar  e_3 = 3\, (\bar c_4 + \bar c_5)\,,
\label{result:large-Nc-chi}
\end{eqnarray} 
valid at NNLO in the expansion. Assuming the approximate validity of (\ref{result:large-Nc-chi}) at $\mu = M_{[8]}$, it suffices
to determine the five parameters $\bar c_4,\bar c_5, \bar c_6$ and $\bar e_1, \bar e_4$. Analogous relations can be used for the
wave-function renormalization terms. As in \cite{Semke:2011ez} we apply the following sum rules
 \begin{eqnarray}
\bar \zeta_D + \bar \zeta_F= \frac 13 \,\bar \xi^{}_D\,, \qquad \bar \zeta_0 =\bar  \xi_0 -\frac{1}{3}\,\bar \xi_D \,.
\label{result:large-Nc-zeta}
\end{eqnarray} 
The parameters $\bar M_{[8]} $ and $\bar M_{[10]} $ are the renormalized and scale-independent bare masses of the baryon 
octet and decuplet. 
They do not coincide with the chiral SU(3) limit of the baryon masses. The latter are determined by a set of nonlinear 
and coupled equations
\allowdisplaybreaks[1]
 \begin{eqnarray}
&& M\ = \bar M_{[8]}   -\, \frac{5 \,C^2  }{768 \,\pi^2\, f^2}\,\frac{\Delta^3\,(2\,M+\Delta)^3}{M^2\,(M+\Delta)^2}\,
\Bigg\{M+ \Delta
\nonumber\\
&& \qquad \qquad +\, \frac{2\,M\,(M+\Delta)+ \Delta^2}{2\,M}  \Bigg\} \,\log \frac{\Delta^2\,(2\,M+ \Delta )^2}{(M+\Delta)^4} \,,
\nonumber\\
&& M + \Delta =  \bar M_{[10]} -\, \frac{C^2  }{384 \,\pi^2\, f^2}\,\frac{\Delta^3\,(2\,M+\Delta)^3}{(M+\Delta)^4}\,
\Bigg\{M
\nonumber\\
&& \qquad \qquad +\,  \frac{2\,M\,(M+\Delta)+ \Delta^2}{2\,(M+\Delta)}  \Bigg\} \,\log \frac{M^4}{\Delta^2\,(2\,M+\Delta)^2}\,,
\label{decompose-bare-masses}
\end{eqnarray} 
where we identified the baryon octet and decuplet masses in the one-loop self energy with
$M$ and $M+ \Delta$ respectively. The parameter $C$ characterizes the octet decuplet transition
coupling strength to a Goldstone boson \cite{Semke2005}. We will return to its numerical value below.

The loop contributions of the baryon self energies were computed before in the infinite volume limit \cite{Semke2005,Semke:2011ez}.
Within the $\chi \overline{MS}$ scheme it suffices to compute, besides the tadpole integral, a subtracted scalar bubble-loop
integral $\bar I_{ Q R}(M_B)$, which depends on a meson mass $m_Q$ of type $Q \in [8]=\{\pi, K, \eta\}$ and baryon masses
$M_{R,B}$ of type $R,B\in [8] = \{N, \Lambda, \Sigma , \Xi\}$
or $R,B \in [10] =\{ \Delta, \Sigma^*, \Xi^*, \Omega\}$. It takes the form
\begin{widetext}
\begin{eqnarray}
&& \bar I_{Q R}(M_B)=\frac{1}{16\,\pi^2}
\left\{ \left(\frac{1}{2}\,\frac{m_Q^2+M_R^2}{m_Q^2-M_R^2}
-\frac{m_Q^2-M_R^2}{2\,M_B^2}
\right)
\,\log \left( \frac{m_Q^2}{M_R^2}\right)
\right.
\nonumber\\
&& \;\quad \;\,+\left.
\frac{p_{Q R}}{M_B}\,
\left( \log \left(1-\frac{M_B^2-2\,p_{Q R}\,M_B}{m_Q^2+M_R^2} \right)
-\log \left(1-\frac{M_B^2+2\,p_{Q R}\,M_B}{m_Q^2+M_R^2} \right)\right)
\right\}
\nonumber\\
&& \;\quad \;\,+\, \Delta \bar I_{QR}(M_B) \, ,
\nonumber\\
&& p_{Q R}^2 =
\frac{M_B^2}{4}-\frac{M_R^2+m_Q^2}{2}+\frac{(M_R^2-m_Q^2)^2}{4\,M_B^2}  \,,
\label{def-bubble}
\end{eqnarray} 
\end{widetext}
where the term $\Delta \bar I_{QR}(M_B)$ vanishes in the infinite volume limit. Here the computation of the
finite volume correction requires special care \cite{Luscher:1986pf,Geng:2011wq,Bali:2012qs,Leskovec:2012gb}. Depending on the specific values
of the meson and baryon masses, power-law contributions in $1/L$ arise. This is contrasted to the exponential behaviour of
the Bessel function in the tadpole integral (\ref{def-tadpole}). It is necessary to discriminate two
cases here. For subthreshold conditions with $M_B < m_Q+ M_R$ we find
 \begin{eqnarray}
&& \Delta \bar I_{QR} =
\frac{1}{8\,\pi^2}\sum^{\vec n \neq 0}_{\vec n \in  Z^3}
\Big( \int_0^1 d z K_0(|x_n| \mu (z)) - \frac{2m_Q \, K_1(|\vec x_n|m_Q)}{|x_n|\,(M_R^2- m_Q^2)} \Big)\,,
\nonumber\\
&& \mu^2(z) = z\,M_R^2+(1-z)\,m_Q^2- (1-z)\,z\,M_B^2 \;, \qquad \vec x_n = L \,\vec n \,,
\label{DeltaIQRnormal}
\end{eqnarray} 
with finite volume effects exponentially suppressed. L\"uscher's power-law terms \cite{Luscher:1986pf} arise for $M_B > M_R+m_Q$. In this case we find three distinct contributions
 \begin{eqnarray}
&& \Delta \bar I_{QR} =  - i\,\frac{p_{QR}}{8\,\pi\,M_B} + \frac{1}{8\,\pi^2\,L\,M_B}\,
Z_{00}\Big(1, \frac{L^2}{4\,\pi^2}\,p^2_{QR} \Big)
\nonumber\\
&& \qquad \;\; \,+\, \frac{1}{2\,\pi^2} \,\Re  \sum^{n \neq 0}_{n \in  Z^3} \int_{0}^{+\infty } d \lambda  \,
\frac{\lambda}{|x_n| }\,e^{-\lambda\,|x_n|/\sqrt{2}}\,e^{+ i\,\lambda\,|x_n|/\sqrt{2}}\,f(  \lambda^2 )  \,.
\label{DeltaIQR-Luescher}
\end{eqnarray} 
The first contribution in (\ref{DeltaIQR-Luescher}) exactly cancels the imaginary part
of (\ref{def-bubble}) that arises for $M_B > M_R+m_Q$. The second term is given by L\"uscher's zetafunction
 \begin{eqnarray}
&& Z_{00}(1, k^2) =\sqrt{\pi}^3 \sum^{|n| \neq 0}_{n \in Z^3}
\int_0^1 \frac{dt}{\sqrt{t}^3} e^{+t\,k^2- 4 \pi^2\,\vec n^2 /t}
+ \sum^{ |\vec n| \neq |k |}_{\vec n \in Z^3} \frac{e^{-\vec n^2+k^2}}{\vec n^2-k^2}
 \nonumber\\
&& \qquad \qquad +\,\sqrt{\pi}^3\,  \Big\{ - 2+
\int_0^1 \frac{dt}{\sqrt{t}^3}\,\Big( e^{+t\,k^2} -1 \Big) \Big\}\,,
\end{eqnarray} 
which we rewrote into a form that is particularly suitable for a numerical evaluation \cite{Leskovec:2012gb}.
Finally, the third term specifies further exponentially suppressed contributions, which we expressed in terms of the function
 \begin{eqnarray}
&& f(\lambda^2) = \frac{1}{2}\,\frac{1}{E_Q\,E_R\,( E_Q+E_R) }\,\frac{M_B^2}{(E_Q+E_R)^2-M_B^2}\,,
\nonumber\\
&& E_Q= \sqrt{m_Q^2+i\,\lambda^2}\,, \qquad \qquad E_R = \sqrt{M_R^2+i\,\lambda^2} \,.
\end{eqnarray} 
We point out that in a strict chiral expansion of the bubble loop (\ref{def-bubble}) the baryon masses $M_B$ and $M_R$ are
expanded  around $M$. This implies that the condition $M_B > M_R + m_Q$ is never met. Therefore L\"uschers
finite power-law terms do not arise in any of the computations \cite{Ren:2012aj,Ren:2013dzt,Ren:2013oaa}.
This is contrasted by the lattice results that indeed satisfy the condition $M_B > M_R + m_Q$  for various cases in the decuplet sector. It is an important property of our self consistent scheme, in which this dominant and crucial effect
is incorporated correctly. Note that even in a case with $M_B < M_R + m_Q$ a chiral expansion of the first term in (\ref{DeltaIQRnormal}) does not converge.

Before providing explicit expressions for the baryon self energies in a finite box, we address a slight complication.
The Passarino-Veltman reduction \cite{Passarino:1978jh}, on which the $\chi \overline{MS}$ scheme rests, requires an 
additional loop integral, which is needed to express any one-loop diagram in terms of a basis of scalar loop integrals. 
Since in a finite box Lorentz invariance is lost the reduction formalism of  Passarino and Veltman \cite{Passarino:1978jh} 
requires an adaptation. This is readily achieved in the case where the three momentum of the baryon is set to zero. The 
evaluation of the baryon self energy leads to integrals of the type
\begin{eqnarray}
&& I^{(a)}_{QR}(p_0) = -\,\frac{i}{V}\,\sum_{\vec n \in Z^3 } \int \frac{d k_0}{(2\pi)}\,\frac{k_0^a}{k^2-m_Q^2}\,\frac{1}{k^2 +p_0^2 -2\,p_0\,k_0 -M_R^2}\,,
\nonumber\\
&& I^{(a)}_{Q}  = \frac{i}{V}\,\sum_{\vec n \in Z^3} \int \frac{d k_0}{(2\pi)}\,\frac{k_0^a}{k^2-m_Q^2} \qquad {\rm with} \qquad 
\vec k = \frac{2\,\pi}{L}\, \vec n \,,
\label{def-generalization-PV}
\end{eqnarray}
where we are interested in their finite volume effects only
\begin{eqnarray}
&& \Delta I^{(a)}_{QR}(p_0) = -\,i\,\sum_{\vec n \in Z^3 }^{\vec n \neq 0}  \int \frac{d^4 k}{(2\pi)^4}\,\frac{e^{i\, \vec x_n \cdot \vec k}\,k_0^a}{k^2-m_Q^2}\,\frac{1}{k^2 +p_0^2 -2\,p_0\,k_0 -M_R^2}\,,
\nonumber\\
&&  \Delta I^{(a)}_{Q}  =i\,\sum_{\vec n \in Z^3}^{\vec n \neq 0} \int \frac{d^4 k}{(2\pi)^4}\,\frac{e^{i\, \vec x_n \cdot \vec k}\,k_0^a}{k^2-m_Q^2} \qquad {\rm with} \qquad 
\vec x_n = L\, \vec n \,.
\label{def-generalization-PV-Delta}
\end{eqnarray}
In contrast to the divergent integrals (\ref{def-generalization-PV}) the integrals (\ref{def-generalization-PV-Delta}) are 
always finite and well behaved. The renormalization of the divergences was considered in great detail in \cite{Semke2005}.
Note that additional structures proportional to $(k_0^2- \vec k^2)^a$ need not to be considered 
separately since it is justified to use the replacement $(k_0^2- \vec k^2)^a \to m_Q^{2\,a}$ in the integrals (\ref{def-generalization-PV-Delta}). 
This is quite immediate for the tadpole type integral $\Delta I^{(a)}_{Q} $ but not so obvious for 
the $\Delta I^{(a)}_{QR}(p_0) $ term. The substitution would be justified rigorously if the integrals of the form
\begin{eqnarray}
-\,i\,\sum_{\vec n \in Z^3 }^{\vec n \neq 0}  \int \frac{d^4 k}{(2\pi)^4}\,\frac{e^{i\, \vec x_n \cdot \vec k}\,k_0^a}{k^2 +p_0^2 -2\,p_0\,k_0 -M_R^2}
=-\,i\,\sum_{\vec n \in Z^3 }^{\vec n \neq 0}  \int \frac{d^4 k}{(2\pi)^4}\,\frac{e^{i\, \vec x_n \cdot \vec k}\,(k_0+p_0)^a}{k^2  - M_R^2}\,,
\end{eqnarray}
would vanish. In fact they do not vanish, however, they are suppressed exponentially with $e^{- M_R\,L}$ and 
therefore can be neglected safely. If one wishes to keep track of such terms it is useful to observe that they lead to 
integrals of the form $\Delta I^{(a)}_{Q}$ with the replacement $m_Q \to M_R$.

It remains to study the generic integrals $\Delta I^{(a)}_{QR}(p_0) $ and $\Delta I^{(a)}_{Q}$. We will show that 
they can be expressed in terms of a specific subset of integrals only. This generalizes the results of 
Passareno and Veltman for our particular case. Using the identity
\begin{eqnarray}
2\,k_0 \,p_0 = \big(k^2-m_Q^2\big ) - \big( k^2 + p_0^2 -2\,p_0\,k_0 -M_R^2\Big) +m_Q^2 +p_0^2-M_R^2\,,
\end{eqnarray}
it follows
\begin{eqnarray}
&& \Delta I^{(a)}_{QR}(p_0) = \left( \frac{m_Q^2 +p_0^2-M_R^2}{2\,p_0} \right)^n \,\Delta I^{(0)}_{QR}(p_0)
\nonumber\\
&& \qquad\qquad \;\, +\, \frac{1}{2\,p_0}\,\sum_{k=0}^{a-1}  \left( \frac{m_Q^2 +p_0^2-M_R^2}{2\,p_0} \right)^k 
\Delta I^{(n-k -1)}_Q + \cdots \,,
\label{IQR-reduction}
\end{eqnarray}
where we neglect any baryonic tadpole contribution that are exponentially suppressed with $e^{-M_R\,L}$. 
The result (\ref{IQR-reduction}) shows that in fact there it is sufficient  to consider $I^{(a)}_{QR}(p_0)$ with 
$a=0$ only. It remains to study the tadpole type integrals $I^{(a)}_{Q}$. We first note that 
$I^{(a)}_{Q} =0$ for $a$ odd. Secondly we observe that in the infinite volume limit all such integrals can be expressed in 
terms of the renormalized tadpole integral $\bar I_Q$ encountered already in (\ref{def-tadpole}). Finally we report that an 
explicit computation reveals that in our case at hand there are contributions from $\Delta I^{(0)}_{Q} $ and 
$\Delta I^{(2)}_{Q} $ only. Therefore altogether 
it suffices to consider three master loop integrals, $\bar I_{QR}(p_0)$, $\bar I_Q$ as studied above 
in (\ref{def-bubble}, \ref{def-tadpole}) together with the additional independent structure 
\begin{eqnarray}
\bar I^{(2)}_Q = \frac{1}{4}\frac{m_Q^4}{(4\pi)^2}
\log \left( \frac{m_Q^2}{\mu^2}\right) - \frac{1}{4\pi^2}\sum^{\vec n \neq 0}_{\vec n \in  Z^3} \frac{m^2_Q}{|\vec x_n|^2}K_2(m_Q|\vec x_n|).
\label{def-I2Q}
\end{eqnarray} 
A comparison of (\ref{def-I2Q}) with (\ref{def-tadpole}) reveals indeed that in the infinite volume limit we have
$m_Q^2\,\bar I_Q \to 4 \,\bar I^{(2)}_Q$.

We are now ready to display the baryon self energies in a finite box as implied by the $\chi \overline{MS}$ scheme.
They are expressed in terms of the loop integrals $\bar I_Q$, $\bar I_{QR}$ and $\bar I_Q^{(2)}$.  We find
\allowdisplaybreaks[1]
\begin{eqnarray}
&&\Sigma^{\rm loop}_{B \in [8]} = \sum_{Q\in [8], R\in [8]}
\left(\frac{G_{QR}^{(B)}}{2\,f} \right)^2  \Bigg\{
\frac{M_R^2-M_B^2}{2\,M_B}\, \bar I_Q
- \frac{(M_B+M_R)^2}{E_R+M_R}\, p^2_{QR}\,
\Big(\bar I_{QR} + \frac{\bar I_Q}{M_R^2-m_Q^2}\,\Big)
\Bigg\}
\nonumber \\
&& \qquad  \;\,\,\,+\sum_{Q\in [8], R\in [10]}
\left(\frac{G_{QR}^{(B)}}{2\,f} \right)^2 \, \Bigg\{
\frac{1}{3}\,\frac{M_B}{M_R^2}\,\bar I^{(2)}_Q
 - \frac{2}{3}\,\frac{M_B^2}{M_R^2}\,\big(E_R+M_R\big)\,p_{QR}^{\,2}\,
\Big(\bar I_{QR} + \frac{\bar I_Q}{M_R^2-m_Q^2}\Big)
\nonumber\\
&& \qquad \qquad
+ \,\Bigg(  \frac{(M_R-M_B)\,(M_R+M_B)^3+m_Q^4}{12\,M_B\,M^2_R}\,
+ \frac{4\,M_B^2+6\,M_R\,M_B-2\,M_R^2}{12\,M_B\,M_R^2}\,m_Q^2\Bigg)\,\bar I_Q
   \Bigg\}
\nonumber\\
&& \qquad  \;\,\,\,+ \,
\frac{1}{(2\,f)^2}\sum_{Q\in [8]} \Bigg(  \Big[ G^{(\chi )}_{BQ} - m_Q^2\,G^{(S)}_{BQ} \Big]\, \bar I_Q - M_B \,G^{(V)}_{BQ}\, \bar I^{(2)}_Q\Bigg) \,,
\label{result-loop-8}
\end{eqnarray} 
with $E_R = \sqrt{M_R^2+p^2_{QR} }$ and
\begin{eqnarray}
&&\Sigma^{\rm loop}_{B\in [10]} = \sum_{Q\in [8], R\in [8]}
\left(\frac{G_{QR}^{(B)}}{2\,f} \right)^2  \Bigg\{
-\frac{1}{3}\,\big( E_R +M_R\big)\,p_{QR}^{\,2}\,
\Big(\bar I_{QR}+ \frac{\bar I_Q}{M_R^2-m_Q^2}\Big)+ \frac{1}{6\,M_B}\,\bar I^{(2)}_Q
\nonumber\\
&& \qquad \qquad
+ \,\Bigg( \frac{(M_R-M_B)\,(M_R+M_B)^3 + m_Q^4}{24\,M^3_B}
- \frac{4\,M_B^2+2\,M_R\,M_B+2\,M_R^2}{24\,M^3_B}\,m_Q^2\Bigg)\,\bar I_Q \Bigg\}
\nonumber\\
&& \qquad \;\,\,\,+\sum_{Q\in [8], R\in [10]}
\left(\frac{G_{QR}^{(B)}}{2\,f} \right)^2 \, \Bigg\{
 -\frac{(M_B+M_R)^2}{9\,M_R^2}\,\frac{2\,E_R\,(E_R-M_R)+5\,M_R^2}{E_R+M_R}\,
p_{QR}^{\,2}\,\Big(\bar I_{QR}
\nonumber\\
&&\qquad \qquad
+ \,\frac{\bar I_Q}{M_R^2-m_Q^2}\Big) + \frac{1}{9\,M_R^2\,M_B} \,\Big[2\,M_B\,M_R+M_R^2-M_B^2 \Big]\,\bar I^{(2)}_Q
\nonumber\\
&&\qquad \qquad +\,\Bigg( \frac{M_R^4+M_B^4+12\,M_R^2\,M_B^2-2\,M_R\,M_B\,(M_B^2+M_R^2)}{36\,M^3_B\,M^2_R}\,
(M^2_R-M^2_B)
\nonumber\\
&&\qquad \qquad  +\,\frac{(M_B+M_R)^2\,m_Q^4}{36\,M_B^3\,M_R^2}
+\frac{4\,M_B^4-4\,M^3_B\,M_R+2\,M_B^2\,M_R^2-2\,M_R^4}{36\,M_B^3\,M_R^2}
\,m_Q^2\Bigg)\,\bar I_Q \Bigg\}
\nonumber\\
&& \qquad  \;\,\,\,+\,
\frac{1}{(2\,f)^2}\sum_{Q\in [8]} \Bigg(  \Big[G^{(\chi )}_{BQ} - m_Q^2\,G^{(S)}_{BQ}\Big] \, \bar I_Q
- M_B \,G^{(V)}_{BQ}\, \bar I^{(2)}_Q \Bigg)\,,
\label{result-loop-10}
\end{eqnarray}
where the sums in (\ref{result-loop-8}, \ref{result-loop-10}) extend over the intermediate Goldstone bosons ($Q\in[8]=\{\pi, \eta, K\}$), the baryon octet ($R\in [8]= \{N, \Lambda, \Sigma, \Xi\}$) and decuplet  states ($R\in[10]= \{\Delta, \Sigma^*, \Xi^*, \Omega \}$).
In the infinite volume limit the self energies coincide with the previous expression
as derived first in \cite{Semke2005,Semke:2011ez}. Upon a systematic expansion of the baryon masses in the quark masses we generate the results of strict chiral perturbation theory. The running of the low-energy constants as described in detail in the Appendix guarantees the independence of the baryon masses on the ultraviolet renormalization scale at N$^3$LO.

The coupling constants $G_{QR}^{(B)}$  are determined by the parameters $F,D,C$ and $H$ that characterize the strength of the meson-baryon 3-point vertices in the chiral Lagrangian. Explicit expressions are listed in \cite{Semke2005}. The parameters $F$ and $D$ follow from a study of semi-leptonic decays of baryon octet states, $B\rightarrow B^\prime + e + \bar \nu_e$. This leads to  $F\simeq 0.45 $ and $D \simeq 0.80$  (see e.g.\cite{Lutz:1999vc}), the values used in this work. Our value of $C\simeq 1.6$ is determined by the hadronic decays of the decuplet baryons (see e.g. \cite{Semke2005}). The parameter $H$ is poorly determined by experimental data so far. Using large-$N_c$ sum rules, the parameters $C$ and $H$ are estimated in terms of the empirical values for $F$ and $D$ \cite{Dashen1994}. It holds
\begin{eqnarray}
H= 9\,F-3\,D \,,\qquad \qquad C=2\,D \,,
\label{large-Nc-HC}
\end{eqnarray}
at subleading order in the $1/N_c$ expansion.

The coupling constants $G_{QR}^{(\chi)}$ probe the renormalized symmetry breaking parameters
$\bar b_0,\bar b_D, \bar b_F, \bar d_0$ and $\bar d_D$, which entered already the tree-level
contribution in (\ref{def-tree-level}). They are detailed in Table I of \cite{Semke:2011ez}.
We are left with 17 symmetry conserving low-energy constants that determine the coupling constants
$G_{QR}^{(S)}$ and $G_{QR}^{(V)}$. They do not depend on the renormalization scale and
are also listed in Table I of \cite{Semke:2011ez}. The large number of unknown chiral parameters at this order is reduced by matching the low-energy and the $1/N_c$ expansions of the product of two axial-vector quark currents \cite{Lutz2002a,LutzSemke2010}.
The 17 parameters are correlated by the 12 sum rules
\begin{eqnarray}
&&g^{(S)}_F = g^{(S)}_0 - \frac 12\, g^{(S)}_1, \qquad h_1^{(S)}=0, \qquad h_2^{(S)}=0\,, \qquad
 h^{(S)}_3 = \frac 32\, g^{(S)}_0 - \frac 94\, g^{(S)}_1 + \frac 12\, g^{(S)}_D\,, \qquad
\nonumber \\
&& h^{(S)}_4 = 3\, \big(g^{(S)}_D+\frac 32\, g^{(S)}_1 \big)\,, \qquad
h^{(S)}_5 = g^{(S)}_D + 3\, g^{(S)}_1, \qquad h^{(S)}_6 = -3\, \big(g^{(S)}_D+\frac 32\, g^{(S)}_1 \big)\,,
\nonumber\\
&&g^{(V)}_D = -\frac 32\, g^{(V)}_1, \qquad g^{(V)}_F = g^{(V)}_0 - \frac 12\, g^{(V)}_1\,,\qquad
h^{(V)}_1 =0\,, \qquad h^{(V)}_2 = \frac 32\, g^{(V)}_0 - 3\, g^{(V)}_1\,, \qquad
\nonumber \\
&& h^{(V)}_3 = \frac 32\, g^{(V)}_1\,,
\label{result:Q2-Nc-constraints}
\end{eqnarray} 
leaving only the five unknown parameters, $g_0^{(S)}, g_1^{(S)}, g_D^{(S)}$ and $g_0^{(V)}, g_1^{(V)}$.

We summarize the number of unknown parameters relevant in the baryon octet and decuplet sector.
At N$^3$LO there are altogether 20 parameters if we apply the large-$N_c$
relations (\ref{result:large-Nc-chi}, \ref{result:large-Nc-zeta}, \ref{result:Q2-Nc-constraints}).


\section{Low-energy parameters from lattice data}
\begin{figure}[t]
\center{\includegraphics[keepaspectratio,width=1.\textwidth]{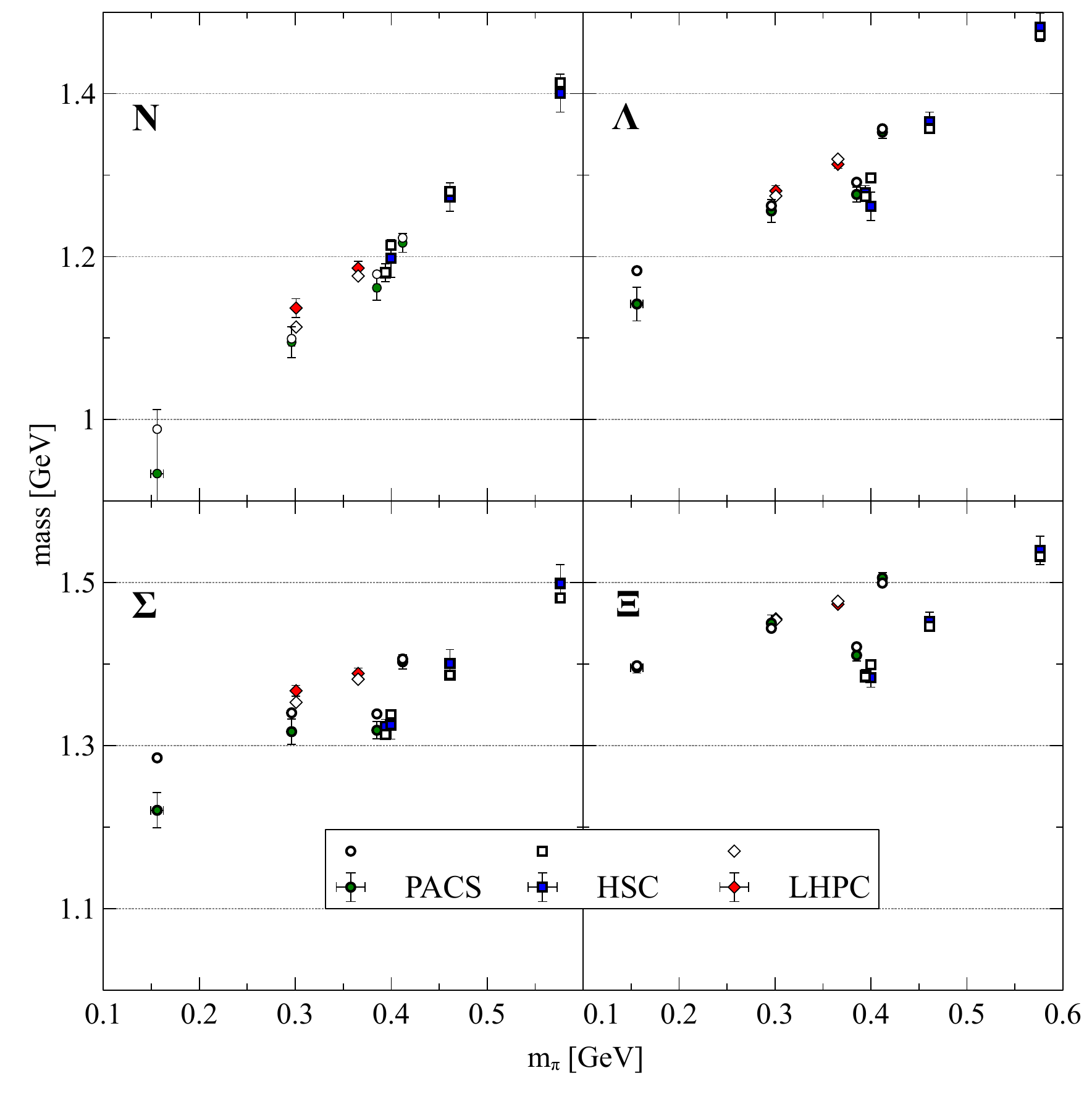}%
} %
\caption{\label{fig:1a} Baryon octet masses compared with lattice data from PACS-CS, HSC and LHPC.
The open symbols give the result of our global fit with $\chi^2/N \simeq 1.73$, $0.73$, $1.93$ respectively.   }
\end{figure}

\begin{figure}[t]
\center{\includegraphics[keepaspectratio,width=1.\textwidth]{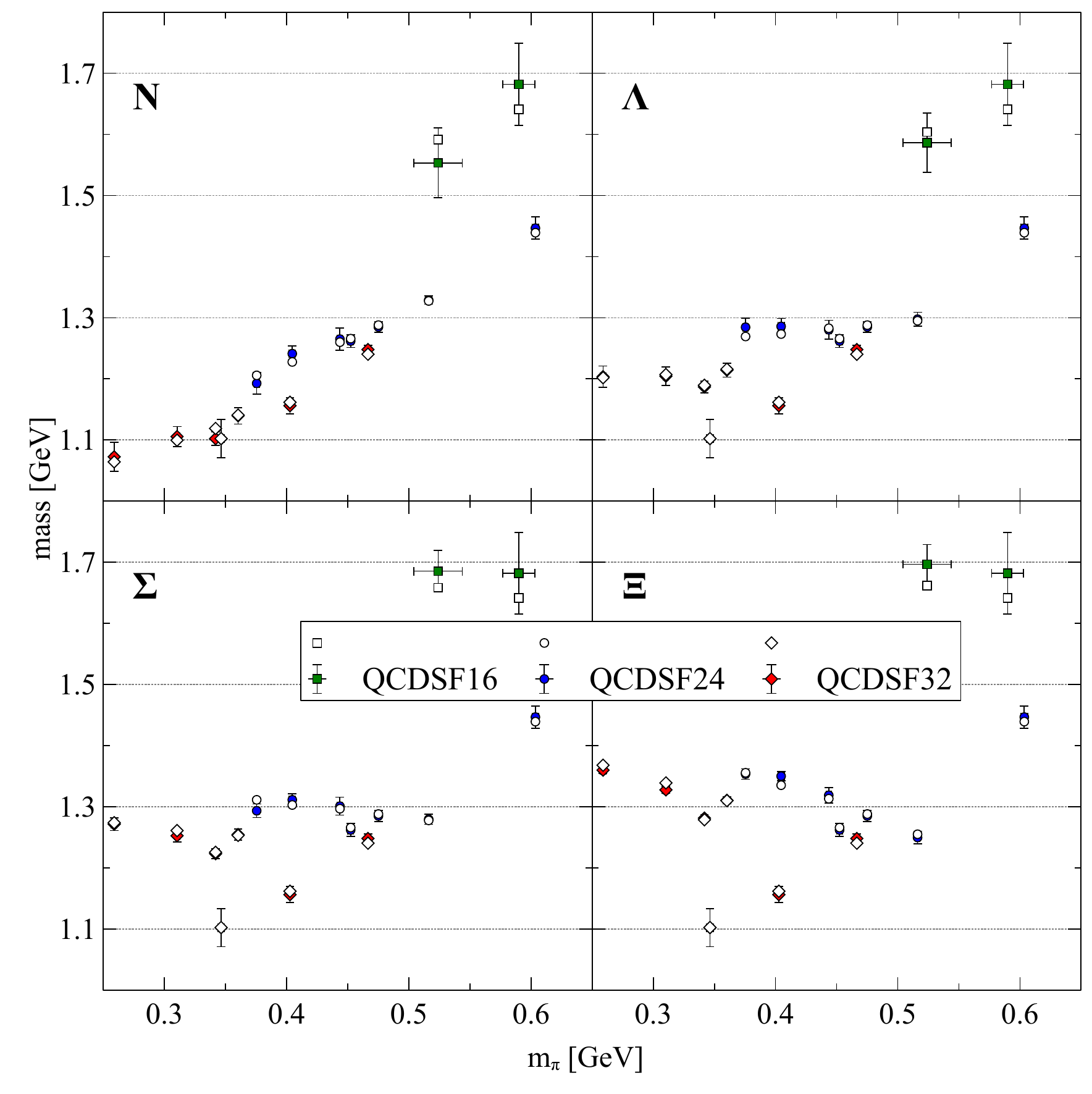}%
}  %
\caption{\label{fig:1b} Baryon octet masses compared with lattice data from QCDSF-UKQCD. The open symbols
give the result of our global fit with $\chi^2/N \simeq 0.48$, $0.48$, $0.56$ for the $16^3$, $24^3$, $32^3$ lattices respectively. }
\end{figure}

\begin{figure}[t]
\center{\includegraphics[keepaspectratio,width=0.50\textwidth]{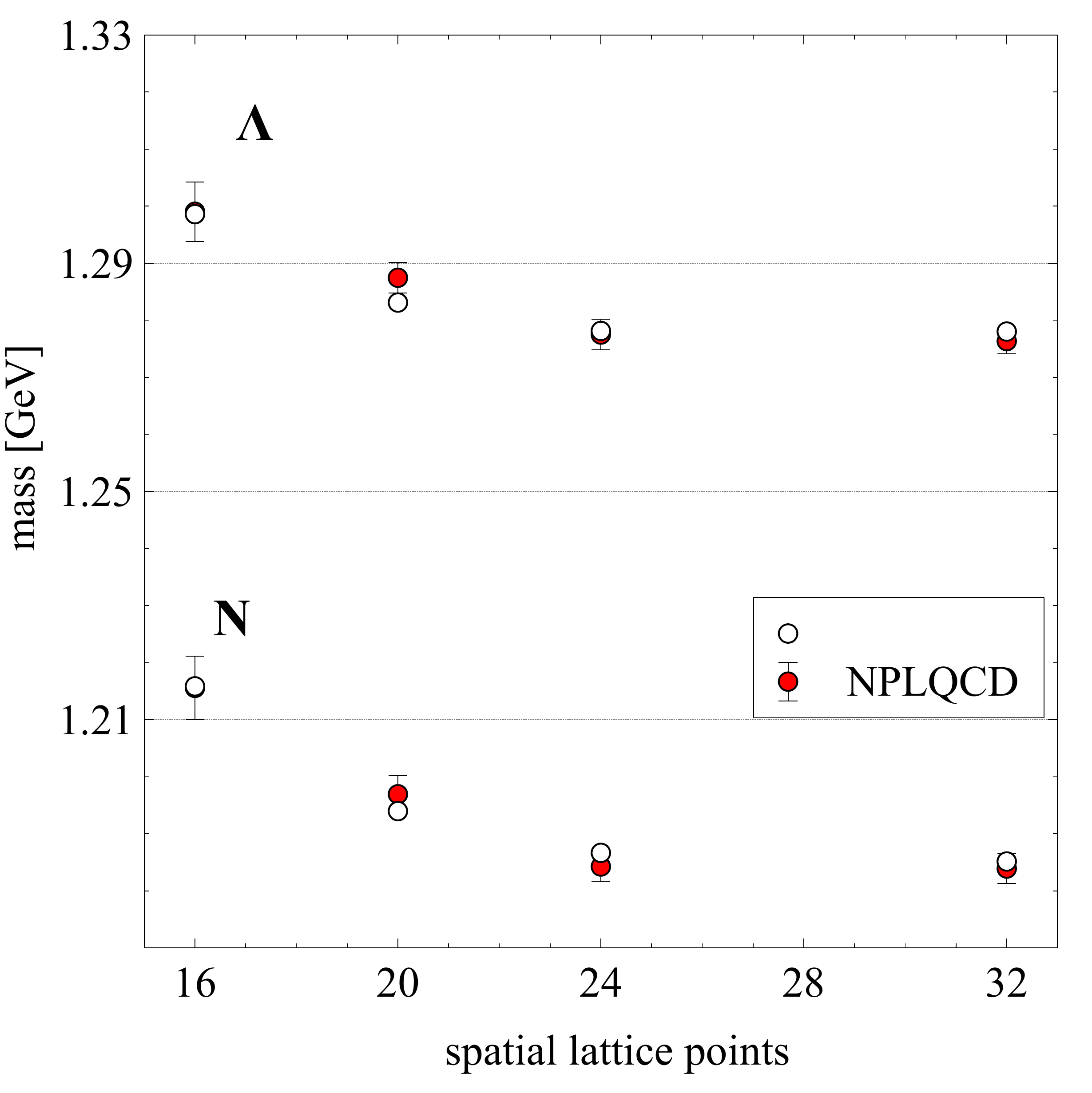}%
\includegraphics[keepaspectratio,width=0.50\textwidth]{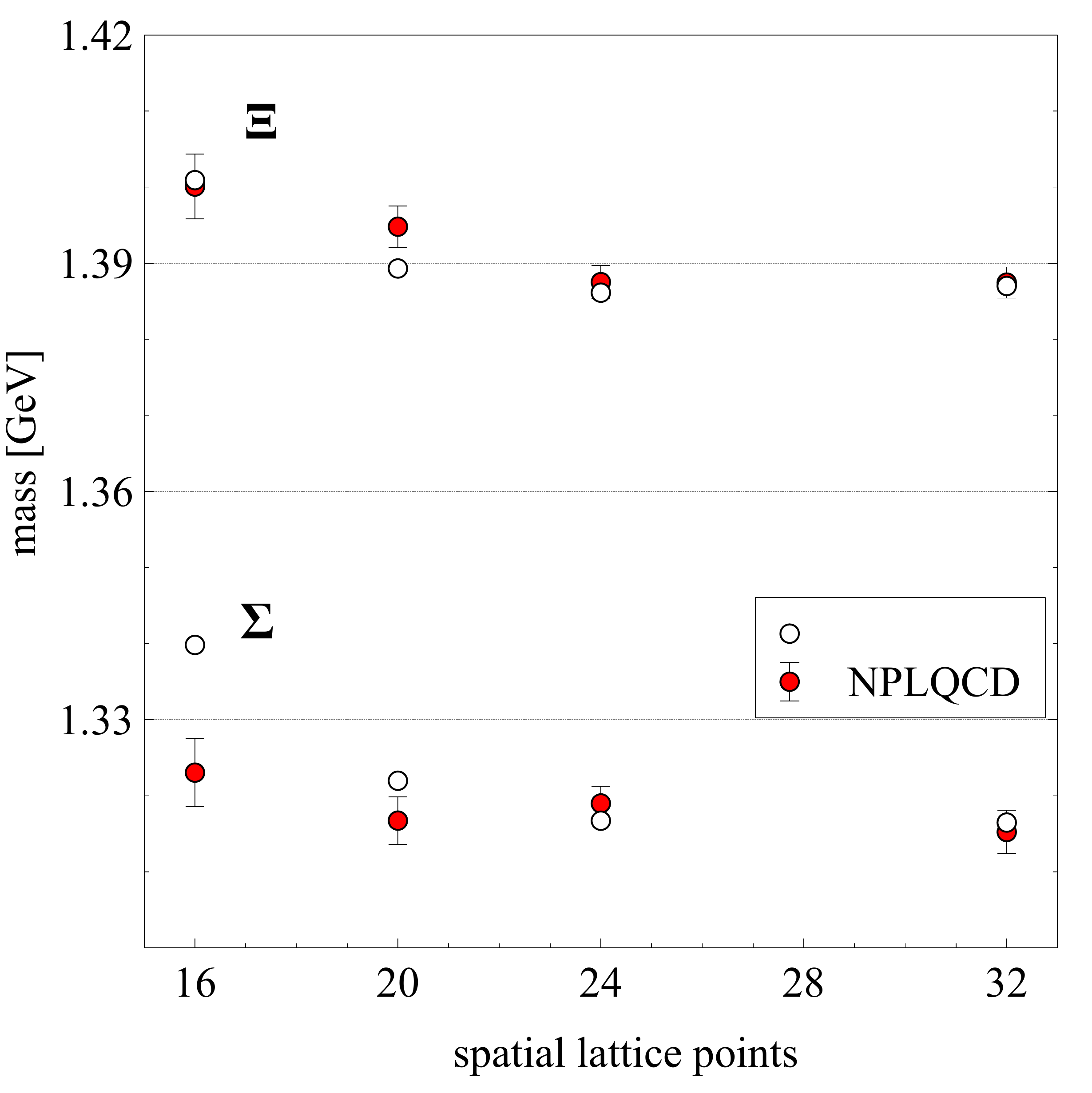}%
}  %
\caption{\label{fig:2}Baryon octet masses compared with lattice data from NPLQCD. The open symbols
give the result of our global fit with $\chi^2/N \simeq 1.75$.   }
\end{figure}

\begin{figure}[t]
\center{\includegraphics[keepaspectratio,width=1.\textwidth]{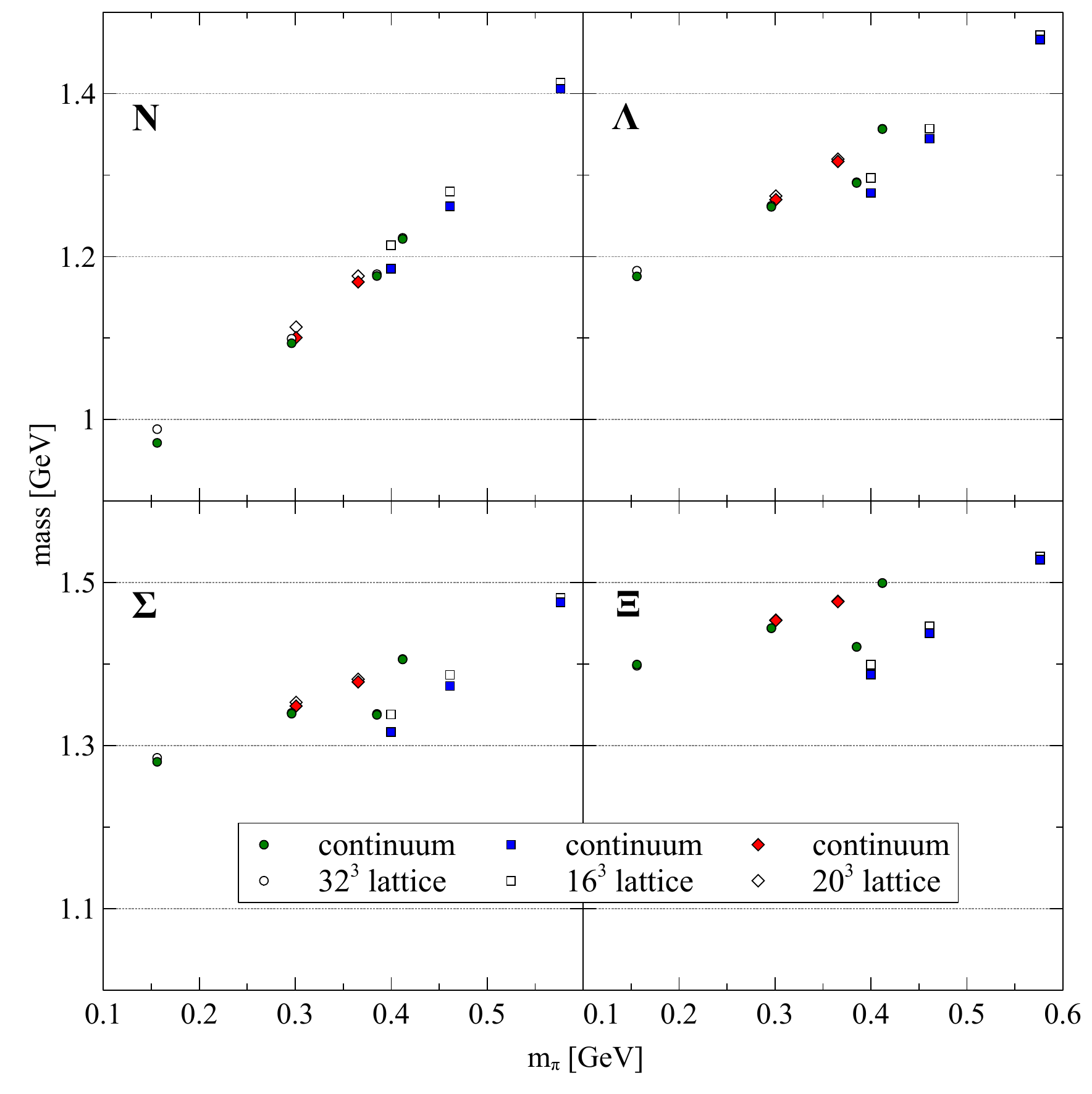}%
} %
\caption{\label{fig:1c} The size of finite volume effects for the baryon octet masses, where the parameter set 1 is used. 
The open symbols are recalled from Fig. \ref{fig:1a} and show  
the results including finite volume effects, the corresponding solid points show the masses in the infinite volume limit.  }
\end{figure}

\begin{figure}[t]
\center{\includegraphics[keepaspectratio,width=1.\textwidth]{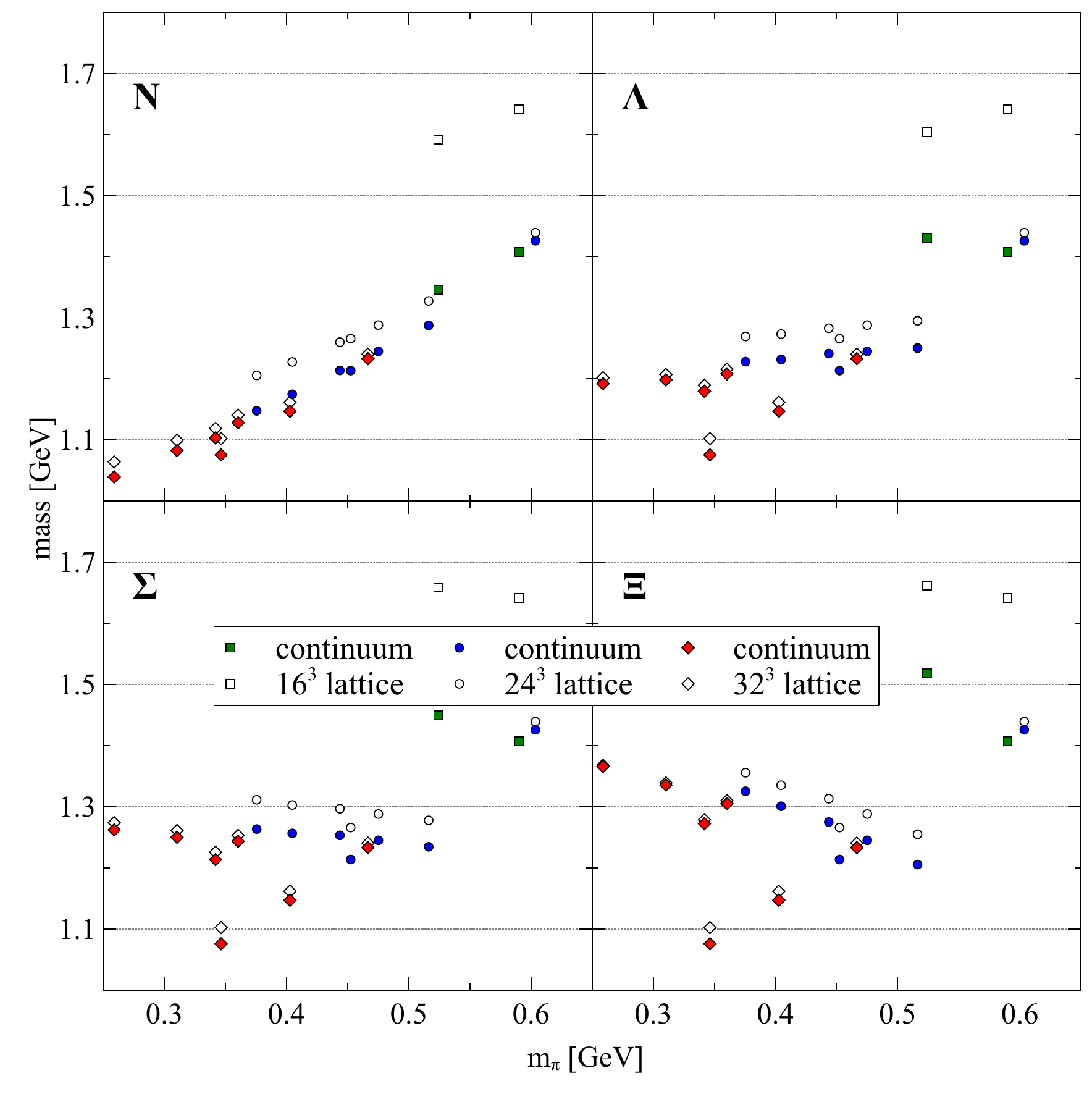}%
}  %
\caption{\label{fig:1d} The size of finite volume effects for the baryon octet masses, where the parameter set 1 is used. 
The open symbols are recalled from Fig. \ref{fig:1b} and show 
the results including finite volume effects, the corresponding solid points show the masses in the infinite volume limit. }
\end{figure}

Our primary goal is the reliable extraction of the low-energy constants from the current lattice data set on the baryon masses. Therefore our analyses rely on the empirical and very precise values for the baryon octet and decuplet masses. Since we do not consider isospin breaking effects nor electromagnetic corrections in our work, we adjust the subset of eight parameters
 \begin{eqnarray}
\bar b_0\;, \bar b_D\;,  \bar b_F\;,  \bar d_0\;,  \bar d_D\;, \bar e_1\;, \bar \zeta_D\; , \bar \xi_D \,,
\label{def-subset}
\end{eqnarray} 
to the isospin averaged baryon masses. The particular subset (\ref{def-subset}) is well suited for that purpose, since for given values of the remaining low-energy constants the parameters are determined by the solution of a linear system.

The use of the empirical baryon masses as suggested in \cite{Semke:2011ez,Semke:2012gs,Lutz:2012mq}
constitutes a significant simplification. Rather than a fit with 20 free parameters the adjustment of only 12 parameters to the lattice data is required. Here one should recall that for each parameter set the determination of the baryon masses requires the solution of a set of eight non-linear equations that are coupled to each other. We note that the works \cite{Ren:2012aj,Ren:2013dzt,Ren:2013oaa} do not use the physical masses in the loop contribution of the baryon self energy and therefore can not always describe finite volume
effects in a reliable and controlled  manner.

In the previous works \cite{Semke:2011ez,Semke:2012gs,Lutz:2012mq} only a subset of the 12 parameters were adjusted to a subset of available lattice data. Data points at small volumes were not considered since the finite volume corrections were not implemented yet. As emphasized before though the subset of lattice data at large volumes can be recovered quite accurately, in such a scheme not all low-energy constants can be determined reliably. Therefore in the scenario of \cite{Semke:2011ez,Semke:2012gs,Lutz:2012mq} all the symmetry preserving parameters that enter at N$^3$LO were put to zero. That left 6 free parameters only that were adjusted successfully to the data of 5 different lattice groups. Moreover it turned out that one parameter, $\xi_0$, could not be determined since its impact on the chi-square function was negligible. So in turn, with in fact only 5 relevant parameters the large volume lattice data
were reproduced with a typical $\chi^2/N \sim 1 - 2$. A remarkable success of the chiral extrapolation that illustrates the consistency of the lattice data from 5 different groups.

In the current work we attempt to determine the full set of low-energy parameters and fit
the 12 parameters to the available lattice data set. We form the chi-square
 \begin{eqnarray}
\chi^2 = \sum_{i=1}^N \left( \frac{M^{\rm Lattice}_i- M_i^{\rm EFT}}{\Delta M_i} \right)^2\,,
\label{def-chi-square}
\end{eqnarray} 
where we take the error $(\Delta M_i)^2$ to be the squared sum of statistical and systematic errors
 \begin{eqnarray}
\Big( \Delta M_i\Big)^2 = \left( \Delta M^{\rm statistical}_i \right)^2 + \left( \Delta M^{\rm systematic}_i\right) ^2\,.
\end{eqnarray} 
We recall that the latter assumption is justified only if the two types of errors are uncorrelated. While the statistical error can be taken from the respective lattice group, the systematic error is poorly known. There are several sources for the latter. Since the available lattice data are evaluated at one lattice spacing only there is an error due to the continuum limit extrapolation. Furthermore, there is an error due to the extraction of the baryon masses from the asymptotic behavior of their correlation function. One would expect the error to be
systematically larger for the decuplet as compared to the octet masses. Finally, there must also be a residual uncertainty from the chiral extrapolation at N$^3$LO.

Additional correlated errors stem from an uncertainty in the determination of the lattice scales of the various groups. Such errors must not be included in the chi-square function via (\ref{def-chi-square}). In previous works \cite{MartinCamalich:2010fp,Geng:2011wq,Ren:2012aj,Ren:2013dzt,Ren:2013oaa} a different chi-square function was formed
 \begin{eqnarray}
\chi^2_{\rm correlated} = \sum_{i,j=1}^N
\left( M^{\rm Lattice}_i- M_i^{\rm EFT} \right) \Big[C^{(-1)}\Big]_{ij}
\left(M^{\rm Lattice}_j- M_j^{\rm EFT} \right)\,,
\label{def-chi-square-correlated}
\end{eqnarray} 
with the correlation matrix
 \begin{eqnarray}
C_{ij} = \delta_{ij}\,\Delta M^{\rm statistical}_i\,\Delta M^{\rm statistical}_i + \Delta M^{\rm scale}_i\,\Delta M^{\rm scale}_j \,,
\label{def-correlation-matrix}
\end{eqnarray} 
being computed in terms of the error $\Delta M^{\rm scale}_j$ that is implied by the scale uncertainty. We would argue that such a treatment is not necessarily always adequate, since it neglects the correlation implied by the quark mass determination via (\ref{meson-masses-q4}). The baryon masses are sensitive to the precise quark masses as obtained from the lattice pion and kaon masses. Clearly, this procedure depends on the lattice scale assumed.
\begin{table}[t]
\setlength{\tabcolsep}{4.5mm}
\renewcommand{\arraystretch}{1.3}
\begin{center}
\begin{tabular}{l||rrrr}\hline
                                          & Fit 1      &  Fit 2      & Fit 3      & Fit 4       \\ \hline \hline
$\bar M_{[8]}\;\;$ \hfill [GeV]           & 0.6370     & 0.6425      & 0.6211     & 0.6171      \\
$\bar M_{[10]}$\hfill [GeV]               & 1.1023     & 1.1129      & 1.1013     & 1.1080      \\

$\bar \xi_0\, \hfill \mathrm{[GeV^{-2}]}$ & 0.9698     & 0.9722      & 0.8284     & 0.8234      \\

$c_4\, \hfill \mathrm{[GeV^{-3}]}$        & -0.1051    & -0.1054     & -0.1157    & -0.0625     \\
$c_5\, \hfill \mathrm{[GeV^{-3}]}$        & -0.0822    & -0.0809     & -0.0590    & -0.0748     \\
$c_6\, \hfill \mathrm{[GeV^{-3}]}$        & -0.7535    & -0.7485     & -0.8542    & -0.9313     \\
$e_4\, \hfill \mathrm{[GeV^{-3}]}$        & -0.3990    & -0.4079     & -0.3948    & -0.4270     \\

$g^{(S)}_0\,\hfill\mathrm{[GeV^{-1}]}$    & -7.5891    & -7.5425     & -7.8378    & -7.9790     \\
$g^{(S)}_1\,\hfill\mathrm{[GeV^{-1}]}$    & 5.9615     & 6.0975      & 6.3602     & 4.8148      \\
$g^{(S)}_D\,\hfill\mathrm{[GeV^{-1}]}$    & -3.4955    & -3.5833     & -3.4694    & -3.0003     \\
$g^{(V)}_0\,\hfill\mathrm{[GeV^{-2}]}$    & 5.5127     & 5.3849      & 3.8422     & 4.2322      \\
$g^{(V)}_1\,\hfill\mathrm{[GeV^{-2}]}$    & -5.2967    & -5.3606     & -6.0160    & -5.3309     \\ \hline

$\chi^2/N$                               & 1.255      &$ \begin{array}{r}1.266 \\1.020  \end{array}  $   & 1.238      & $\begin{array}{r}1.263 \\1.019 \end{array}$
 \\

\hline
\end{tabular}
\caption{The parameters are adjusted to reproduce the baryon octet and decuplet masses
from the PACS-CS, HSC, LHPC, QCDSF-UKQCD and NPLQCD groups as described in the text.
The remaining low-energy parameters are determined by the large-$N_c$
sum rules (\ref{result:large-Nc-chi}, \ref{result:large-Nc-zeta}, \ref{large-Nc-HC}, \ref{result:Q2-Nc-constraints}) together
with the condition that the isospin averaged masses of the physical baryon octet and decuplet states are reproduced
exactly.
}
\label{tab:FitParametersA}
\end{center}
\end{table}

In our scheme we can avoid this issue altogether by including the lattice scales of the various groups as free fit 
parameters. In turn our chi-square function is defined with (\ref{def-chi-square}), where only statistical and 
uncorrelated systematical errors are required. Here we assume that the unknown systematical errors, to which we 
will return below,  are uncorrelated. The constraint, that we always reproduce the empirical baryon masses, allows 
us to do so and perform our own scale setting. In the previous work \cite{Semke:2012gs} it was demonstrated that for 
instance the chi-square for the QCDSF-UKQCD data is a very steep function in the lattice scale. The given uncertainty 
in the lattice scale leads to a large and significant variation of the chi-square, with a typical range of 
$2 < \Delta \chi^2/N  < 10$. It was argued in \cite{Semke:2012gs} that ultimately a controlled chiral extrapolation 
of the baryon masses, indeed may lead to a very precise determination of the lattice scale.

Finally, before starting a parameter search we have to state our assumption for the systematic errors. In the absence of 
a solid estimate of the latter, we will let float their size. First, we will perform fits, where only statistical errors 
are considered. From the quality of such fits we may get a hint about the size of the systematic errors. In a second step
we will redo such fits where we assume a non-vanishing systematic but uncorrelated error in order to lower the optimal 
chi-square down to about $\chi^2 / N \simeq 1$. We will assume universal errors for the octet and decuplet masses. A 
first rough estimate for the expected size follows from the isospin splitting in the baryon masses. Since our scheme and 
also the lattice groups assume perfect isospin symmetry, such an estimate should constitute a lower boundary for the 
systematic errors. The isospin splitting of the baryon octet masses is in the range of about 1-4 MeV. A similar range 
is observed for the decuplet masses. The uncertainty in the low-energy parameters will be deduced from a variation of 
parameters as they come out of the fits with different sizes of the systematic error.

\begin{figure}[t]
\center{\includegraphics[keepaspectratio,width=1.\textwidth]{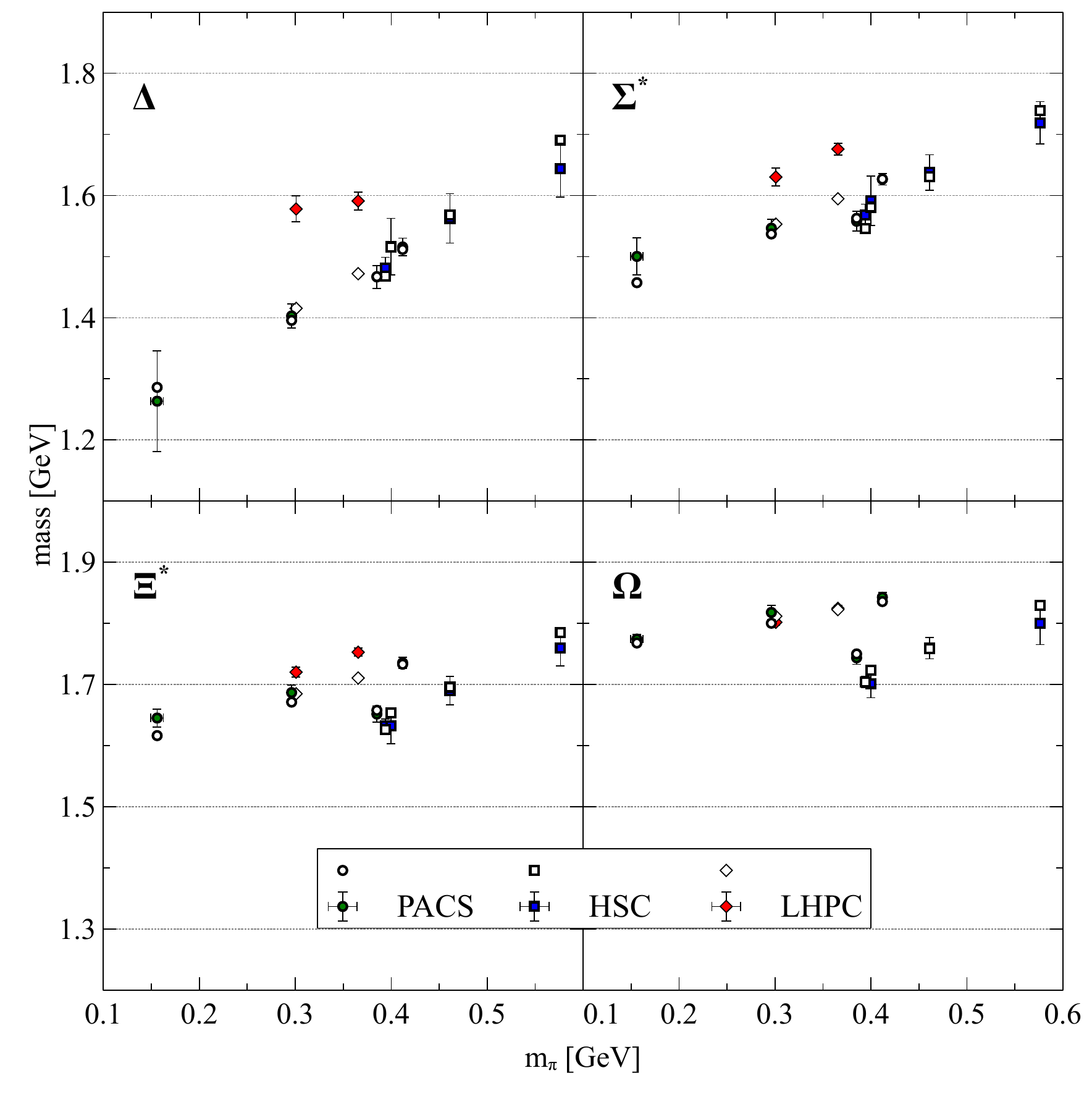}%
}  %
\caption{\label{fig:3a} Baryon decuplet masses compared with lattice data from PACS-CS, HSC and LHPC. The open symbols
give the result of our global fit with $\chi^2/N \simeq 0.80$, $0.44$, $35.7$ respectively.
The $\Delta, \Sigma^*$ and $\Xi^*$ masses of the LHP collaboration were not included in our global fits. The chi-square
given in Tab. \ref{tab:FitParametersA} does not consider the contributions of those outliers.}
\end{figure}

\begin{figure}[t]
\center{\includegraphics[keepaspectratio,width=1.\textwidth]{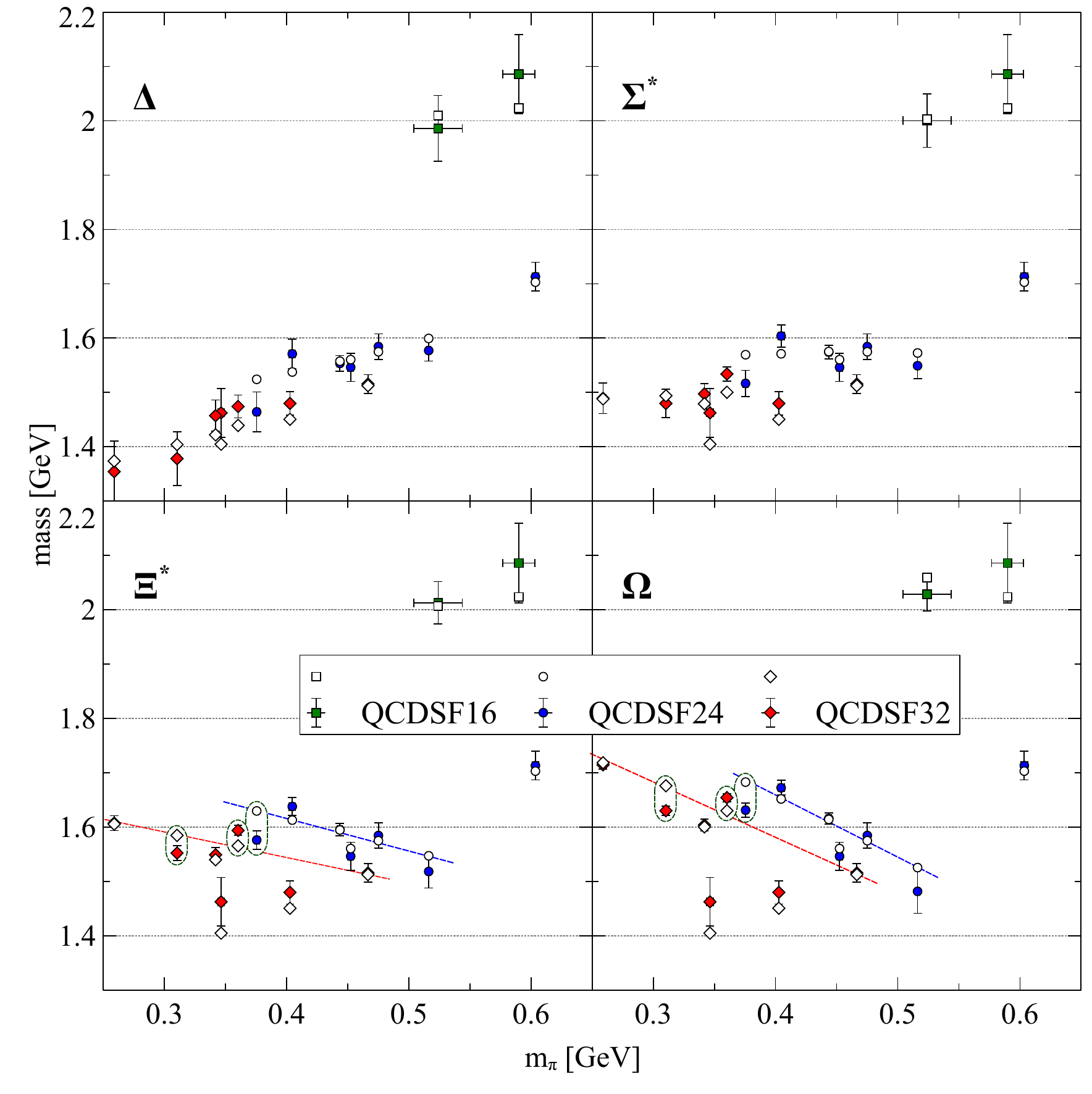}%
}  %
\caption{\label{fig:3b} Baryon decuplet masses compared with lattice data from QCDSF-UKQCD. The open symbols
give the result of our fit with $\chi^2/N \simeq 0.51$, $1.75$, $2.87$ for the $16^3$, $24^3$, $32^3$ lattices
respectively. Along the straight lines the averaged quark mass $2\,m+ m_s$ is approximatively constant.
The quark mass ratio $(m_s-m)/(2\,m +m_s)$ takes the values shown in (\ref{ratio:32}) and (\ref{ratio:24}). }
\end{figure}

We take all data points from PACS-CS, HSC, LHPC, QCDSF-UKQCD and NPLQCD into account with a pion mass smaller than about 
600 MeV. In the Figs. \ref{fig:1a}- \ref{fig:3b} the lattice data are shown in physical units using already the optimal 
lattice scales as they come out of our fit. We chose a representation linear in the pion  mass as to highlight 
the approximate linearity of the baryon masses in that 
parameter \cite{LHPC2008,WalkerLoud:2008pj,Walker-Loud:2013yua,Lutz:2013kq}. The scattering of the different lattice 
points reflects different choices for the kaon masses. Our chi-square functions exclude the data points of the BMW 
collaboration since they are not made available publicly. A comparison with their nucleon and omega mass as displayed 
in a figure of \cite{BMW2008} for their smallest lattice spacing will be provided below nevertheless. Finally we exclude 
all data points from LHPC on the decuplet states, with the exception of the omega baryon. We found no way to reproduce 
the $\Delta, \Sigma^*$ and $\Xi^*$ masses in any of our fits and therefore consider them as outliers. The consequence of 
our best fit will be confronted with those data points also.

Stable results for the low-energy constants require a simultaneous fit of all data points included in our chi-square 
function. Leaving out for instance the octet data from LHPC leaves a rather flat valley in the chi-square function along 
which the low-energy parameters may change significantly. A determination of the symmetry conserving parameters relevant at 
N$^3$LO  is most sensitive to the QCDSF-UKQCD data on their $16^3$, $24^3$ and $32^3$ lattices and the
NPLQCD data for the baryon octet masses on their $16^3$, $20^3$, $24^3$ and $32^3$ lattices. Since the PACS-CS and HSC data 
are for quite distinct values of the pion and kaon masses, as compared to QCDSF-UKQCD and NPLQCD, it is crucial to keep 
their data in the chi-square function also.
\begin{table}[t]
\setlength{\tabcolsep}{3.5mm}
\renewcommand{\arraystretch}{1.2}
\begin{center}
\begin{tabular}{l||cccc|cl} \hline
                                                &  Fit 1   &  Fit 2    &  Fit 3    & Fit 4    & Lattice Result    \\ \hline
$a_{\rm QCDSF}\,  \hfill \mathrm{[fm]}$         &  0.07389 & 0.07405   &  0.07368  & 0.07412  & 0.0765(15)  \hfill \cite{Bietenholz:2011qq} \\
$a_{\rm HSC} \,   \hfill \mathrm{[fm]}$         &  0.03410 & 0.03417   &  0.03405  & 0.03423  & 0.0351(2)   \hfill \cite{HSC2008} \\
$a_{\rm LHPC}\,   \hfill \mathrm{[fm]}$         &  0.12080 & 0.12125   &  0.12072  & 0.12122  & 0.1241(25)  \hfill \cite{WalkerLoud:2008bp} \\
$a_{\rm PACS}\,   \hfill \mathrm{[fm]}$         &  0.09055 & 0.09061   &  0.09045  & 0.09053  & 0.0907(14)  \hfill \cite{PACS-CS2008} \\

\\

$f\,  \hfill \mathrm{[MeV]}$         &  92.4      &  92.4      &  89.5       &  88.1    &   \\
$10^3\,(L_4 -2\,L_6)\, $             &  0.099     &  0.099     &  0.110      &  0.092   &   \\
$10^3\,(L_5 -2\,L_8)\, $             & -0.392     & -0.392     & -0.393      & -0.357   &   \\
$F$                                  &  0.450     &  0.450     &  0.436      &  0.427   &   \\
$D$                                  &  0.800     &  0.800     &  0.802      &  0.785   &   \\
\hline

\end{tabular}
\caption{Our determination of the lattice scale for QCDSF-UKQCD, HSC, LHPC and PASC.
The values of the additional five low-energy parameters that were activated in Fit 3 and Fit 4 are shown.
}
\label{tab:FitParametersB}
\end{center}
\end{table}

\begin{figure}[t]
\center{\includegraphics[keepaspectratio,width=1.\textwidth]{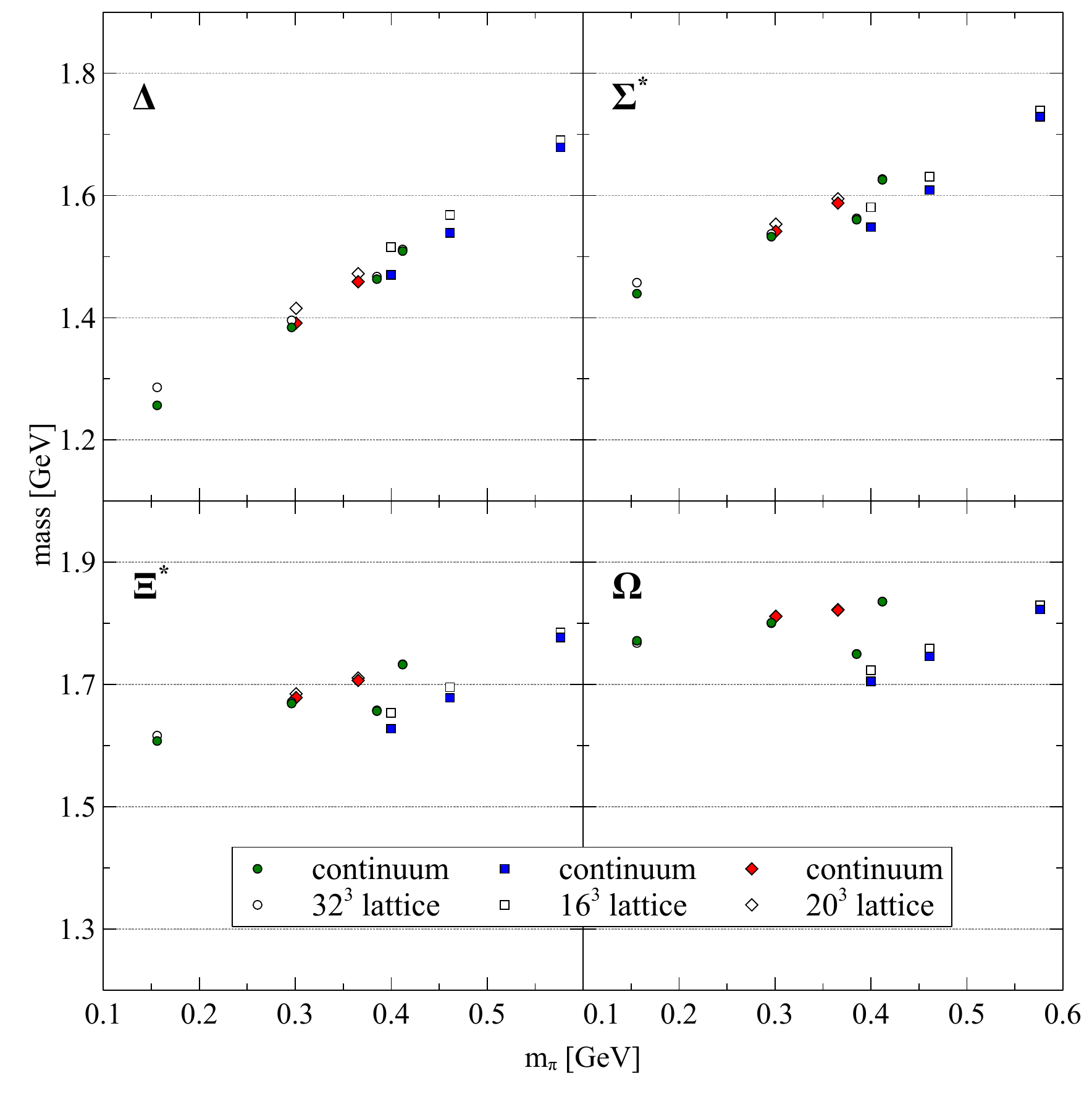}%
} %
\caption{\label{fig:3c} The size of finite volume effects for the baryon decuplet masses, where the parameter set 1 is used. 
The open symbols are recalled from Fig. \ref{fig:3a} and show  
the results including finite volume effects, the corresponding solid points show the masses in the infinite volume limit.    }
\end{figure}

\begin{figure}[t]
\center{\includegraphics[keepaspectratio,width=1.\textwidth]{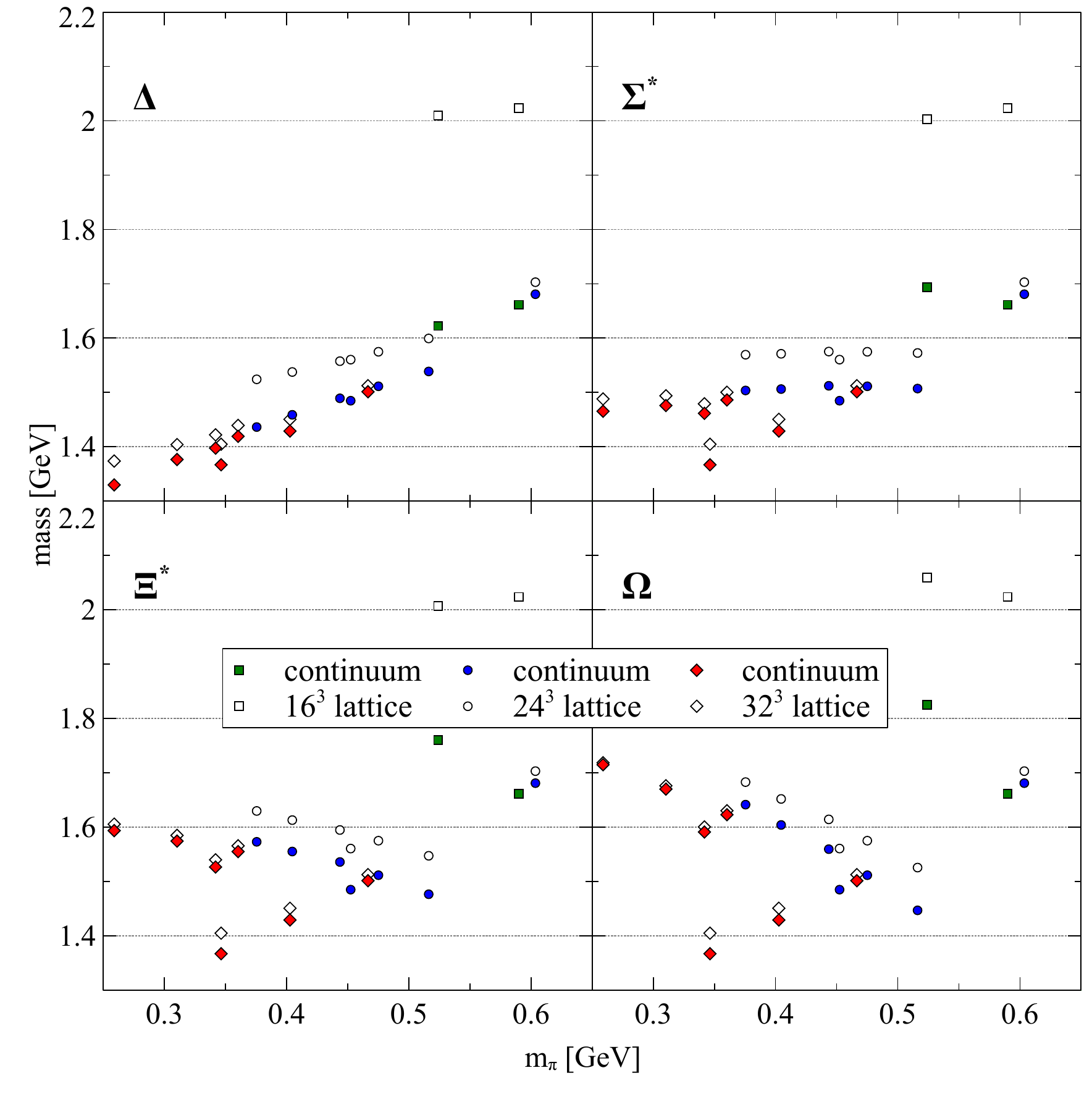}%
}  %
\caption{\label{fig:3d} The size of finite volume effects for the baryon decuplet masses, where the parameter set 1 is used. 
The open symbols are recalled from Fig. \ref{fig:3b} and show  
the results including finite volume effects, the corresponding solid points show the masses in the infinite volume limit.  }
\end{figure}

In the Figs. \ref{fig:1a}-\ref{fig:3d} the results of our best fit are shown and confronted with the data from the 
various lattice groups. An excellent reproduction is achieved with 12 free parameters only. The total chi-square per number 
of data point is $\chi^2/N \simeq 1.255 $. The corresponding low-energy parameters correspond to Fit 1 in Tab. \ref{tab:FitParametersA}. 
We do not provide any statistical error estimate since they are negligible. This is because we describe more than 220 data 
points with 12 parameters only. The stability of our solution is studied by assuming an additional systematic error for the 
octet and decuplet masses of 3  and 6 MeV respectively. A new parameter search was performed, which results are shown under 
Fit 2 in Tab. \ref{tab:FitParametersA}. While the $\chi^2/N \simeq 1.020$ is lowered significantly all low-energy parameters 
have moderate changes only. Taking the Fit 2 parameters we calculate the chi square value in absence of the systematic 
error. In this case we obtain $\chi^2/N \simeq 1.266$, a value slightly larger than the one of Fit 1. With scenario 3 and 4 
in Tab. \ref{tab:FitParametersA} and Tab. \ref{tab:FitParametersB} we illustrate the influence of including 5 extra 
parameters $F,D$ and $f, L_4-2\,L_6,L_5-2\,L_8$  in the parameter search. While Fit 3 corresponds to the assumption of 
vanishing systematic errors, the Fit 4 parameters are with respect to the same systematic errors as assumed for Fit 2. 
Again the low-energy parameters shown in Tab.\ref{tab:FitParametersB} suffer from moderate variations only.
The chi-square for Fit 4 that includes the additional systematic error is $\chi^2/N \simeq 1.019$. The values of the 
additional five parameters are collected in Tab. \ref{tab:FitParametersB}. We observe that the consideration of the extra 
five parameters does not lower our chi-square value significantly. Our results show that our a priori choices were very 
reasonable. It is pointed out that the values of the low-energy constants $f, L_4 -2\,L_6, L_5 -2\,L_8, F, D$ have a crucial 
impact on the description of the baryon masses from lattice QCD simulations. For instance if we insisted on a positive value 
for $L_5 -2\,L_8$ we did not find fits of comparable quality. The relevance of physical choices for the $F$ and $D$ 
parameters was emphasized previously in \cite{WalkerLoud:2011ab,Semke:2012gs}.

The chiral limit values of the baryon octet and decuplet states, $M$ and $M+ \Delta$, follow from the solution of the set 
of  non-linear equations (\ref{def-non-linear-system}). Using the parameters of Tab. \ref{tab:FitParametersA} and 
Tab. \ref{tab:FitParametersB} we find the ranges
\begin{eqnarray}
M =( 802 \pm 4 )\,{\rm MeV} \,, \qquad \quad
M+ \Delta  = (1103 \pm 6)\, {\rm MeV} \,.
\end{eqnarray}
Our result is compatible with the previous estimates in \cite{Semke:2012gs}. Using parameter set 1 of that work we find 
the values $M \simeq 826$ MeV and $M+ \Delta \simeq 947$ MeV. The somewhat larger values $M \simeq 903$ MeV and 
$M+ \Delta \simeq 1157$ MeV follow with  parameter set 2 of that work.

We remind the reader that our chi-square definition does consider statistical errors as given by the lattice groups only. 
The uncertainty from the scale setting is considered by keeping the various lattice scales as free parameters. 
In Tab. \ref{tab:FitParametersB} we show the lattice scales as they come out of our four different fits. Note that the 
lattice scales of the NPLQCD and HSC groups coincide. It is reassuring that we obtain values that are compatible with the 
scale determination of the lattice groups within their estimated uncertainties. We find remarkable that the variance of 
the lattice scales found for our 4 different fits is at the five per mil level.

In Figs. \ref{fig:1a}-\ref{fig:2} and Figs. \ref{fig:3a}-\ref{fig:3b} a detailed comparison of our Fit 1 results with the 
lattice data is offered. We affirm that any of the other scenarios would lead to almost indistinguishable figures. In 
Fig. \ref{fig:1a} we confront our results for the baryon octet masses with those of PACS-CS, HSC and LHPC. From our global 
fit we obtain $\chi^2/N \simeq 1.73, 0.73, 1.93$ respectively, where the chi-square is evaluated according 
to (\ref{def-chi-square}). Only statistical errors are considered. The uncertainties in the lattice scales were taken into 
account by including the latter into the parameter search. If we instead compute the chi-square with respect 
to (\ref{def-chi-square-correlated}) as 
advocated in \cite{MartinCamalich:2010fp,Geng:2011wq,Ren:2012aj,Ren:2013dzt,Ren:2013oaa} we obtain significantly smaller 
values $\chi^2/N \simeq 1.34, 0.57, 1.69$. From this we conclude that the use of the correlation matrix in the form of 
(\ref{def-chi-square-correlated}) is not always adequate.

In Fig. \ref{fig:1b} we collected the data from the QCDSF-UKQCD collaboration for the baryon octet masses at three different 
lattice volumes. Our global description implies chi-square values $\chi^2/N \simeq 0.48, 0.48, 0.56$ with respect to 
(\ref{def-chi-square}) and $\chi^2/N \simeq 0.27, 0.31, 0.27$  with respect to (\ref{def-chi-square-correlated}). It is 
noteworthy that the lattice data are well described for all three lattice volumes. This was not possible in the previous 
works  \cite{Ren:2012aj,Ren:2013dzt}, which considered exclusively the four octet masses in terms of fits with their 19 
free parameters.

A further important constraint for the low-energy parameters arises from the accurate data
of the NPLQCD group, which we scrutinize in Fig. \ref{fig:2}.  Our global fit describes the latter with 
a $\chi^2/N \simeq 1.75$ and $\chi^2/N \simeq 1.46$ according to (\ref{def-chi-square}) 
and (\ref{def-chi-square-correlated}) respectively. The chi square is largely dominated by an outlier, 
the sigma baryon mass for the $16^3$ lattice.

We conclude our discussion of the baryon octet states with Fig. \ref{fig:1c} and Fig. \ref{fig:1d} where the importance
of the finite volume effects is illustrated. The two figures correspond to the previous  Fig. \ref{fig:1a} 
and Fig. \ref{fig:1b}, only that a comparison of our finite volume results (open symbols) with their infinite volume 
limits (solid symbols) rather than the lattice data is shown. The figures clearly show the significance of the finite 
volume corrections, in particular for the data from the QCDSF-UKQCD group on their 24$^3$ and 16$^3$ lattice. 
Note that the size of the finite volume effects does not always decrease with increasing pion mass. 
This is because the lattice configurations are such that the associated kaon and eta masses are sometimes 
getting smaller as the pion mass increases. This is the case for the QCDSF-UKQCD data. 

We turn to the decuplet masses. In Fig. \ref{fig:3a} they are compared with the results of PACS-CS, HSC and LHPC. 
The two types of chi-squares are $\chi^2/N \simeq 0.80, 0.44, 35.7$ and $\chi^2/N \simeq 0.34, 0.22, 16.1$. While we 
achieve an excellent description of the PACS-CS and HSC data we fail to recover the $\Delta$, $\Sigma^*$ and $\Xi^*$ 
masses of the LHP collaboration. As mentioned before the latter are considered as outliers and consequently
were excluded in our definition of the global chi-square functions. From Fig. \ref{fig:3a} one may speculate that 
there is some tension in the data of LHPC as compared to PACS-CS and HSC, at least for the $\Delta$ where the 
slightly different strange quark masses used are expected not to be relevant.

In Fig. \ref{fig:3b} we confront our results with the QCDSF-UKQCD simulations for the decuplet masses. We refrain 
from  using a speed plot representation suggested in \cite{Bietenholz:2011qq} and also used in previous 
works  \cite{Lutz:2012mq,Ren:2013oaa}. As already shown in \cite{Lutz:2013kq} not taking the particular ratios 
required by the fan plots reveals interesting and additional information on the baryon decupet states. Our description 
of the lattice is characterized by the following chi-square values $\chi^2/N \simeq 0.51, 1.75, 2.87 $ for the 
$16^3$, $24^3$ and $32^3$ lattices. Again computing the chi-square using the correlation 
matrix (\ref{def-correlation-matrix}) leads to significantly smaller values $\chi^2/N \simeq 0.26, 0.87, 1.41 $. 
Our reproduction of the QCDSF-UKQCD is characterized by two outliers for the $32^3$ lattice and one outlier for 
the $24^3$ lattice. While our description for the more complicated $\Delta$ and $\Sigma^*$ masses is excellent, 
this is not the case for the description of the $\Xi^*$ and $\Omega^*$ masses, for which we encircled the outliers. 
We find this puzzling since in particular for the $\Omega$ there is no available decay channel which may complicate 
the computation of finite volume effects. For the $32^3$ lattice the chi-square is largely dominated by the 2nd point 
for the $\Xi^*$ and $\Omega$ mass. Such points have a large distance to the dashed lines shown in Fig. \ref{fig:3b}. 
The latter lines connect smoothly four distinct open symbols that give the prediction of our extrapolation. Similarly, 
the 1st point of the $24^3$ lattice deviates strongly from the dashed lines drawn for the $\Xi^*$ and $\Omega$ masses. 
In this case the line connects five open symbols. Note that analogous straight lines may be drawn for the $\Delta$ and 
$\Sigma^*$ in Fig. \ref{fig:3b} and also for the baryon octet states in Fig. \ref{fig:1b}.

An explanation for such a linearity is readily found. The corresponding lattice points are taken at constant averaged 
quark masses, where the values for $2\,m + m_s$ are changing by less than 3 per cent on the $32^3$ and less than 1 per 
cent for the $24^3$ lattices. On the other hand the quark mass ratio takes the values
\begin{eqnarray}
\frac{m_s-m}{2\,m+m_s} = \Big\{  +0.67, +0.53, +0.38,  0.00\Big\}\,,
\label{ratio:32}
\end{eqnarray}
along the four particular points on the $32^3$ lattice and
\begin{eqnarray}
\frac{m_s-m}{2\,m+m_s} = \Big\{  +0.36, +0.26, +0.16,  0.00, -0.17\Big\}\,,
\label{ratio:24}
\end{eqnarray}
along the five particular points on the $24^3$ lattice. It is an amusing observation that the step size in the symmetry 
breaking quark mass ratio coincides with the corresponding step size in the pion mass. To this extent the linear behaviour 
seen in the Fig. \ref{fig:3b} is a direct consequence of the well known equal spacing rule for the decuplet masses. An 
expansion of the decuplet masses in the parameter $m_s-m$ at fixed value of $2\,m +m_s$ provides an accurate description 
of the masses already with only the linear term kept. Given our framework there is no way to generate the significant 
departure of the lattice data for the $\Xi^*$ and $\Omega$ masses from the straight dashed lines as shown in 
Fig. \ref{fig:3b}. We conclude that the linearity of the decuplet fan plots \cite{Bietenholz:2011qq} is in part a subtle 
consequence of taking particular ratios.

\begin{table}[t]
\setlength{\tabcolsep}{3.5mm}
\renewcommand{\arraystretch}{1.0}
\begin{center}
\begin{tabular}{l||rrrrr}\hline
                                             &  Fit 625      & Fit 525   & Fit 425A &       Fit 425B   & Fit 425C  \\ \hline \hline
$\bar M_{[8]}\;\;$ \hfill [GeV]              & 0.63413  & 0.59844    & 0.59515 & 0.59945 & 0.62992 \\
$\bar M_{[10]}$\hfill [GeV]                  & 1.09281 & 1.09614 & 1.06418 & 0.99809 & 1.03836 \\

$\bar \xi_0\, \hfill \mathrm{[GeV^{-2}]}$   & 1.01089 & 0.79578 & 0.75710 & 0.90093 & 0.96789 \\

$c_4\, \hfill \mathrm{[GeV^{-3}]}$         & -0.21601 & -0.39418 & -0.33941 & -0.15380 & -0.10026 \\
$c_5\, \hfill \mathrm{[GeV^{-3}]}$            & 0.02621 & 0.27852 & 0.26157 & 0.15654 & -0.04887 \\
$c_6\, \hfill \mathrm{[GeV^{-3}]}$           &  -0.68692 &  -0.37708 &  -0.38501 &  -0.55271 &  -0.70022 \\
$e_4\, \hfill \mathrm{[GeV^{-3}]}$           & -0.38886 & -0.35039 & -0.30906 & -0.22345 & -0.31497 \\

$g^{(S)}_0\,\hfill\mathrm{[GeV^{-1}]}$        &  -9.23670 &  -9.47963 &  -9.33530 &  -8.72593 &  -8.21464  \\
$g^{(S)}_1\,\hfill\mathrm{[GeV^{-1}]}$       & 6.59741 & 7.39081 & 6.89777 & 7.62471 & 5.71282 \\
$g^{(S)}_D\,\hfill\mathrm{[GeV^{-1}]}$        &  -2.22149 &  -1.40226 &  -1.46690 &  -4.90674 &  -3.17883 \\
$g^{(V)}_0\,\hfill\mathrm{[GeV^{-2}]}$     & 6.59077 & 3.59025 & 4.43491 & 7.99845 & 5.74075 \\
$g^{(V)}_1\,\hfill\mathrm{[GeV^{-2}]}$        & -4.53460 & -5.67925 & -4.58849 & -2.90244 & -4.83729 \\ \hline \hline

$a_{\rm QCDSF}\,  \hfill \mathrm{[fm]}$           &  0.07401 & 0.07425 & 0.07412 & 0.07423 & 0.07396 \\
$a_{\rm HSC} \,   \hfill \mathrm{[fm]}$          &  0.03412 & 0.03429 & 0.03430 & 0.03448 & 0.03419 \\
$a_{\rm LHPC}\,   \hfill \mathrm{[fm]}$           &  0.12077 &  0.12129 &  0.12176 &  0.12198 &  0.12140 \\
$a_{\rm PACS}\,   \hfill \mathrm{[fm]}$         & 0.09057 & 0.09094 & 0.09111 & 0.09136 & 0.09091  \\

\hline \hline
$\chi^2/N$                                         & 0.8857 & 0.9259 & 1.1319  & 1.1326  & 1.1764

  \\

$\chi^2/N$ @ $L= \infty $  & 6.6783 & 6.3526 & 3.4259 & 3.6110 & 3.4716 \\

\hline
\end{tabular}
\caption{The parameters are adjusted as for Fit 1 of Tab. \ref{tab:FitParametersC}. In the chi square function
the errors for the encircled 6 points in Fig. \ref{fig:3b} are enlarged  by an ad-hoc factor of 5 as
explained in the text. In addition only subsets of the global lattice data sets are considered. While the first and second column
consider lattice data only with $m_\pi< 625$ MeV and $m_\pi< 525$ MeV, the last three columns correspond to the same pion mass cutoff of
$425$ MeV.
}
\label{tab:FitParametersC}
\end{center}
\end{table}

We conclude our discussion of the baryon decuplet states with Fig. \ref{fig:3c} and Fig. \ref{fig:3d} where the relevance 
of the finite volume effects is illustrated. The two figures correspond to the previous  Fig. \ref{fig:3a} 
and Fig. \ref{fig:3b}, only that a comparison of our finite volume results (open symbols) with their infinite volume 
limits (solid symbols) rather than the lattice data is shown. The figures clearly show the importance of the finite volume 
corrections, in particular for the data from the QCDSF-UKQCD group on their 24$^3$ and 16$^3$ lattice. For the 32$^3$ 
lattice the corrections are found to be most sizeable for the $\Delta$ and $\Sigma^*$ masses.  

As a further consistency check we performed additional fits where the relevance of the possible outliers in the QCDSF-UKQCD 
data set as discussed above is reduced. For the encircled 6 points in  Fig. \ref{fig:3b} we enlarged the error by an ad-hoc 
factor of 5. Based on such a modified chi square function we performed a series of fits where the size of the global data 
set is gradually reduced by imposing a cutoff in the maximum pion mass. In order to gauge such new fits we first evaluate 
the updated chi square values with respect to our established solution 
Fit 1 of Tab. 1. We obtain the following values
\begin{eqnarray}
& \chi^2/N \Big|_{m_\pi <625 \,{\rm MeV}} = 0.9085 \qquad {\rm with} \qquad & N=218 \,{\rm data} \; {\rm points} \,,
\nonumber\\
& \chi^2/N \Big|_{m_\pi <525 \,{\rm MeV}} = 0.9697 \qquad {\rm with} \qquad & N=194 \,{\rm data} \; {\rm points} \,,
\nonumber\\
& \chi^2/N \Big|_{m_\pi <475 \,{\rm MeV}} = 1.0508 \qquad {\rm with} \qquad & N=170 \,{\rm data} \; {\rm points} \,,
\nonumber\\
& \chi^2/N \Big|_{m_\pi <425 \,{\rm MeV}} = 1.2309 \qquad {\rm with} \qquad & N=138 \,{\rm data} \; {\rm points} \,,
\nonumber\\
& \chi^2/N \Big|_{m_\pi <375 \,{\rm MeV}} = 1.2098 \qquad {\rm with} \qquad & N=\phantom{0}66 \,{\rm data} \; {\rm points} \,.
\label{reduced-chisquares}
\end{eqnarray}
As a consequence of the reduced impact of the six outliers the chi square value for the first case with $m_\pi < 625$ MeV is 
smaller than the corresponding value of Tab. 1 of about 1.26. As can be seen from the various figures discussed above our 
lattice data description is quite uniform in quality. This is reflected in an only moderate increase of the chi square 
values as one lowers the pion mass cutoff from 625 MeV down to 375 MeV. Note that this reduces the  size of the global 
data set by about a factor 3. The large step from the 425 MeV case to the 375 MeV case is due to the throw out of the HSC 
and NPLQCD data sets. Since all chi square values are quite close to one the significance of refits of the parameter set 1 
using the updated chi square functions is unclear. In particular, we note that if we supplement the updated chi square 
functions by the ad-hoc systematical errors of 3 and 6 MeV for all octet and decuplet baryon masses, as used before in 
Fit 2, the chi square values of (\ref{reduced-chisquares}) would be 1.02 for the $m_\pi <425$ MeV and 1.04 for the 
$m_\pi <$ 375 MeV case. Nevertheless we show 5 representative refits where the maximum pion mass allowed was fixed to 
either 625 MeV, 525 MeV or 425 MeV. The results thereof are collected in Tab. \ref{tab:FitParametersC}. The last row of 
the table illustrates the crucial importance of finite volume effects. Switching them off leads to a huge increase of the 
chi square value.

\begin{figure}[t]
\center{\includegraphics[keepaspectratio,width=0.5\textwidth]{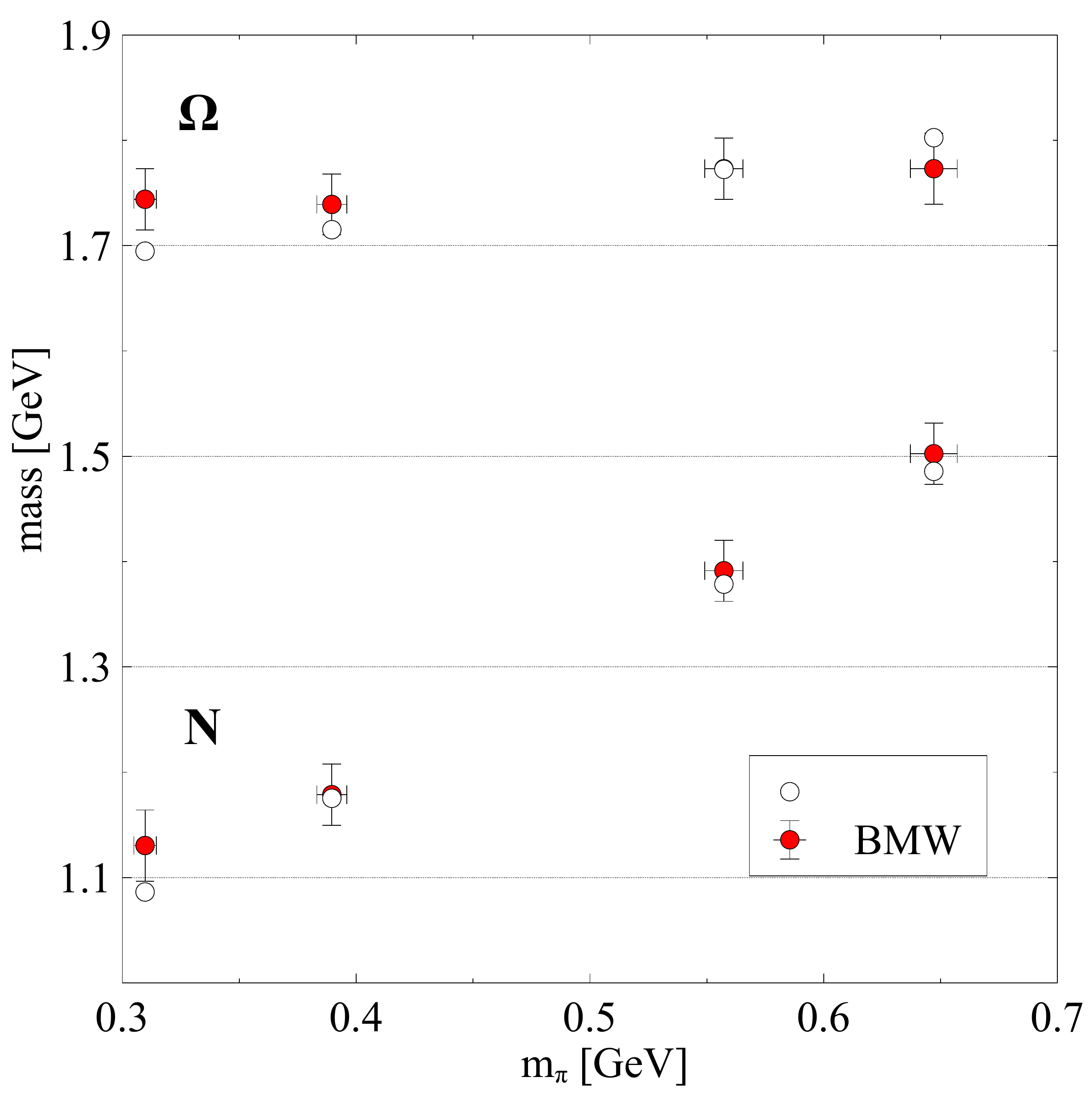}%
}  %
\caption{\label{fig:4}Masses of the nucleon and omega compared with lattice data from BMW. The open symbols
give the prediction of our fit with $\chi^2/N \simeq 0.82 $.  }
\end{figure}

\begin{table}[t]
\setlength{\tabcolsep}{3.5mm}
\renewcommand{\arraystretch}{1.1}
\begin{center}
\begin{tabular}{l||rrrrr}\hline
                                             &  Fit 475A & Fit 475B  & Fit 375A  & Fit 375B & Fit A  \\ \hline \hline
$\bar M_{[8]}\;\;$ \hfill [GeV]              & 0.6300    & 0.9425   & 0.6356   & 0.9495   &  0.6370 \\
$\bar M_{[10]}$\hfill [GeV]                  & 1.1027    & 1.1894   & 1.1066   & 1.2617   &  1.1023 \\

$\bar b_0\, \hfill \mathrm{[GeV^{-1}]}$      & -1.3460   & -0.2979   & -1.3808  & -0.2472  & -1.4118\\
$\bar b_D\, \hfill \mathrm{[GeV^{-1}]}$      &  0.4957   &  0.0673   & 0.4844   &  0.0446  &  0.5022 \\
$\bar b_F\, \hfill \mathrm{[GeV^{-1}]}$      & -0.5859   & -0.1576   & -0.5721  & -0.1437  & -0.5917 \\

$\bar d_0\, \hfill \mathrm{[GeV^{-1}]}$      & -0.2956   & -0.2482   & -0.2789  & -0.1503  & -0.3303 \\
$\bar d_D\, \hfill \mathrm{[GeV^{-1}]}$      & -0.4569   & -0.5441  & -0.4496  & -0.4582  & -0.5196 \\

$F$                                          & 0.4117    & 0.1062 & 0.4138   &  0.1217 &  0.4500  \\
$D$                                          & 0.7863    & -0.1814   & 0.8096   & -0.0386 & 0.8000   \\
\hline \hline
$\chi^2/N$                                   & 4.24      &  3.19       & 4.25     & 1.25    & $\begin{array}{c}
                                                                                              61.0 \\ 40.2
                                                                                              \end{array}$ \\
\hline
\end{tabular}
\caption{The parameters at N$^2$LO are adjusted to the lattice data as in Tab. \ref{tab:FitParametersC}. 
The large-$N_c$ sum rules $C= 2\,D$ and $H = 9\,F-3\,D$ together with $f= 92.4$ MeV and $\mu= M_{[8]}$  are used. 
The fits are performed at fixed lattice scales with $a_{\rm LHPC}  = 0.124$ fm,  $a_{\rm HSC}   = 0.035$ fm,
$a_{\rm PACS}  = 0.091$ fm and $a_{\rm QCDSF} =0.075$ fm.
}
\label{tab:FitParameters:N2LO}
\end{center}
\end{table}

We observe that our refits lead to only rather moderate improvements in the chi square values. This is seen by a comparison 
of (\ref{reduced-chisquares}) with the second last row of Tab. \ref{tab:FitParametersC}. On the other hand the spread in 
the parameter variations is larger as seen in Tab. \ref{tab:FitParametersA} and \ref{tab:FitParametersB}. This is not 
surprising since reducing the size of the lattice data set may not always lead to a full determination of all low-energy 
constants. While we expect the chiral extrapolation to work more accurately at smaller quark masses, it is not clear whether 
a reduced data set with smaller pion masses is sufficient to identify a unique set of parameters. Indeed, our solutions 
425A and 425B have very similar chi square values but significantly different low-energy parameters. This is typical for 
the considered system. Only with the data set as considered in Fit 1-4 of Tab. \ref{tab:FitParametersA} we find reasonably 
well defined and essentially unique solutions. However, even in this case the chi square function is quite shallow around 
the established solution and further lattice data may move the solution somewhat. It is interesting to evaluate the chi 
square of all solutions given in Tab. \ref{tab:FitParametersC} with respect to the 'full' data set. While for the 1st 
column the given chi square value follows, for the last 4 columns we obtain the values 0.90, 0.95, 1.01 and 0.94 
respectively. From those values we may conclude that solution 425A is superior over the solutions 425B and 425C.

As a final consistency check of our lattice data description we provide Fig. \ref{fig:4}, where
our results are compared with the unfitted BMW data as taken from a figure in \cite{BMW2008} for their smallest 
lattice spacing. We use Fit 1 here for which its chi square is estimated with $\chi^2/N \simeq 0.82$. A similar chi square 
is obtained for Fit 2-4. This gives us further confidence on the reliability of our chiral extrapolation. 
Unfortunately, corresponding results for the remaining members of the octet and decuplet states are not available publicly. 
It is interesting to note that the solutions of Tab. \ref{tab:FitParametersC} are discriminated by the BMW data. 
While solution 525A has a chi square of 1.09 solution 425B leads to a significantly worse description with a chi square 
value of 1.32. An even better description with a chi square value of 0.94 is generated by solution 425C. The first two 
solutions have chi square values of 1.05 and 0.86.

There remains an important issue which we have not addressed so far. How much does the quality of the lattice data 
description improve as we increase the accuracy level of the chiral extrapolation formulae? In order to illustrate 
the significance of the N$^3$LO approach we performed various fits to the lattice data at N$^2$LO. A first chiral 
extrapolation study at the self consistent N$^2$LO was reported in \cite{Semke2007} prior to the availability of 
accurate lattice data. Like in Tab. \ref{tab:FitParametersC} we perform fits to the data set where the maximum pion mass 
is restricted by a upper cutoff value. The parameters in Tab. \ref{tab:FitParameters:N2LO}  correspond to 
an upper pion mass of 375 MeV and 475 MeV for the first two and the next two solutions respectively. All 9 parameters that are 
relevant at N$^2$LO are listed in the table. Typically, for a given pion mass cutoff we 
find two solutions of reasonable $\chi^2/N$. The solutions labeled with Fit 375B and Fit 475B are characterized by axial 
coupling constants $F$ and $D$ that are significantly smaller than their known physical values. The solutions 
Fit 375A and Fit 475A have larger chi square values but predict values for $F$ and $D$ that are almost compatible with 
their empirical estimates. In order to further discriminate the two types of solutions we compute the baryon masses for the 
physical quark masses. While for the two A-type solutions the mean deviation from the empirical masses is about 10 MeV 
always, it is significantly larger for the two B-type solutions, for which the mean deviation is about 44 MeV in both cases. 
Note that all four solutions did not consider the physical baryon masses in their chi square definition. 
From this observation we conclude that in fact the A-type solutions are superior to the B-type solutions even though they 
describe the lattice data significantly worse. A further parameter set labeled by Fit A, given in the last column of \ref{tab:FitParameters:N2LO}, 
determined the low-energy parameters $\bar b_0, \bar b_D, \bar b_F$ and $\bar d_0, \bar d_D$ from the physical baryon masses. 
In this case we use the empirical estimates $F=0.45$ and $D=0.80$ together with values for the parameters $\bar M_{[8]}$ and $\bar M_{[10]}$  
from Fit 1 of Tab. \ref{tab:FitParametersA}. This guarantees the fit to be compatible with our previous estimate of the baryon masses in 
their flavour SU(3) limit. The mean deviation of the baryon masses from their empirical values is also about 10 MeV in this case. 
Naturally, the quality of the lattice data description is worse here with $\chi^2/N \simeq 61.0$ and $\chi^2/N \simeq 40.2$ 
for the pion mass cut off at 375 MeV and 475 MeV respectively. Nevertheless, it is comforting to see that the values of the 
low-energy constants vary rather moderately from the solutions Fit 475A, Fit 375A to Fit A. The quality of the 
N$^3$LO lattice data description in Tab. \ref{tab:FitParametersA} is significantly improved as compared to the 
N$^2$LO results of Tab. \ref{tab:FitParameters:N2LO}. This may justify the application of the chiral extrapolation at the 
self consistent N$^3$LO as used throughout this work.

\begin{table}[t]
\setlength{\tabcolsep}{2.5mm}
\renewcommand{\arraystretch}{1.2}
\begin{center}
\begin{tabular}{ll||cccccccc} \hline
 $a\,m_\pi$ & $a \,m_K$ & $a\,m_N$      & $a\,m_\Lambda$    & $ a\,m_\Sigma$  & $a\,m_\Xi$ &
                          $a\,m_\Delta$ & $a\,m_{\Sigma^*}$ & $ a\,m_{\Xi^*}$ & $a\,m_\Omega$  \\ \hline

 0.124, & 0.252    & 0.511  & 0.577 & 0.607 & 0.654 & 0.662 & 0.718 & 0.772 & 0.825  \\
 0.149, & 0.261    & 0.558  & 0.608 & 0.639 & 0.673 & 0.719 & 0.757 & 0.802 & 0.844  \\
 0.145, & 0.259    & 0.539  & 0.595 & 0.624 & 0.664 & 0.693 & 0.739 & 0.787 & 0.834  \\
 0.141, & 0.257    & 0.527  & 0.588 & 0.616 & 0.660 & 0.676 & 0.728 & 0.779 & 0.829  \\
 0.158, & 0.262    & 0.544  & 0.600 & 0.626 & 0.667 & 0.692 & 0.739 & 0.786 & 0.832  \\
 0.173, & 0.267    & 0.565  & 0.614 & 0.638 & 0.674 & 0.714 & 0.754 & 0.797 & 0.838  \\
 0.199, & 0.277    & 0.592  & 0.635 & 0.654 & 0.687 & 0.736 & 0.772 & 0.809 & 0.845  \\
 0.223, & 0.288    & 0.620  & 0.658 & 0.672 & 0.702 & 0.759 & 0.791 & 0.823 & 0.854  \\ \hline

 0.107, & 0.212    & 0.455  & 0.506 & 0.530 & 0.569 & 0.593 & 0.636 & 0.679 & 0.720  \\
 0.126, & 0.218    & 0.471  & 0.518 & 0.540 & 0.575 & 0.608 & 0.647 & 0.686 & 0.725  \\
 0.155, & 0.228    & 0.500  & 0.539 & 0.557 & 0.586 & 0.632 & 0.664 & 0.698 & 0.730  \\
 0.180, & 0.237    & 0.527  & 0.559 & 0.574 & 0.598 & 0.655 & 0.682 & 0.709 & 0.736  \\
 0.194, & 0.244    & 0.549  & 0.577 & 0.588 & 0.610 & 0.676 & 0.699 & 0.723 & 0.746  \\ \hline

 0.070, & 0.169    & 0.339  & 0.389 & 0.411 & 0.446 & 0.442 & 0.485 & 0.526 & 0.565  \\
 0.080, & 0.173    & 0.347  & 0.394 & 0.416 & 0.451 & 0.450 & 0.491 & 0.530 & 0.569  \\
 0.098, & 0.178    & 0.363  & 0.406 & 0.425 & 0.456 & 0.466 & 0.501 & 0.537 & 0.572  \\
 0.121, & 0.176    & 0.391  & 0.418 & 0.433 & 0.453 & 0.495 & 0.517 & 0.541 & 0.564  \\

\end{tabular}
\caption{Our prediction for the baryon octet and decuplet masses in units of the lattice scale $a$.
The pion and kaon masses are taken from the ETM collaboration, where the first, second and third block
in the table corresponds to $\beta $ values of $1.90$, $1.95$ and $2.10$ respectively. The uncertainty
from taking either the parameters of Fit 1-4 is less than 0.002 in all cases.
}
\label{tab:ETMprediction}
\end{center}
\end{table}

Based on the successful description of the currently available lattice data on the baryon masses we find it justified to 
generate detailed predictions for ongoing QCD lattice simulations of the ETM group. First results on the pion and nucleon 
mass for three distinct beta values are reported in \cite{Alexandrou:2013joa} based on 2 +1 +1 simulations with twisted 
mass fermions. As a courtesy of C. Alexandrou we received the corresponding values of the kaon masses, with which we can 
attempt a prediction for their baryon octet and decuplet masses. In the first two columns of Tab. \ref{tab:ETMprediction} 
we recall the lattice pion and kaon mass in units of the lattice scale. Like in our previous analysis we determine the 
optimal lattice scale for the three different beta values using our parameter sets. We find values
\begin{eqnarray}
a^{\beta = 1.90}_{\rm ETM} \simeq 0.0980\, {\rm fm} \,,\qquad
a^{\beta = 1.95}_{\rm ETM} \simeq 0.0880 \,{\rm fm} \,, \qquad a^{\beta = 2.10}_{\rm ETM} \simeq 0.0674 \,{\rm fm} \,,
\label{def-ETM-scale}
\end{eqnarray}
within the range obtained in \cite{Alexandrou:2013joa}. Taking the lattice scales (\ref{def-ETM-scale}) we compute the 
nucleon mass with respect to the ETM specifications and obtain with our Fit 1 parameters a 
$\chi^2/N \simeq 0.65$, $\chi^2/N \simeq 1.89$ and $\chi^2/N \simeq 0.29$ for the three beta values. Very similar 
values are obtained for all solutions of Tab. \ref{tab:FitParametersA} and \ref{tab:FitParametersC}, with the exception 
of solution B that leads to slightly worse chi square values of 1.13, 2.42 and 0.39 respectively. In the last eight 
columns of Tab. \ref{tab:ETMprediction} we give our predictions for the baryon masses, where we take the average of the 
results obtained with our 4 different fit scenarios of Tab. \ref{tab:FitParametersA}. The maximum deviation from our 
average is smaller or equal to $0.002$ for all solutions of Tab. \ref{tab:FitParametersA}. The spread is slightly larger 
if we consider also the solutions of \ref{tab:FitParametersC}. Excluding the disfavored solution 245B the maximum 
uncertainty is $0.004$. The Fit $245B$ leads to a slightly larger spread of $0.007$.

\clearpage

\section{Pion and strangeness baryon sigma terms}

The pion- and strangeness baryon sigma terms are important parameters in various physical systems. They play a crucial role in dark 
matter searches (see e.g. \cite{Belanger:2013oya,Crivellin:2013ipa}). The strangeness sigma term is a key parameter for the determination 
of a possible kaon condensate in dense nuclear matter \cite{Kaplan:1986yq}. The pion-nucleon sigma term is of greatest relevance in the 
determination of the density dependence of the quark condensate at low baryon densities and therefore provides a first estimate of the 
critical baryon density at which chiral symmetry may be restored
(see e.g. \cite{Lutz:1999vc}).

Assuming exact isospin symmetry with $m_u=m_d\equiv m$,  the pion-nucleon sigma term $\sigma_{\pi N}$ and the strangeness sigma term $\sigma_{sN}$ are defined as follows
 \begin{eqnarray}
&&\sigma_{\pi N} =  m\,\frac{\partial}{\partial m} m_N\,, \qquad
 \sigma_{s N} =  m_s\,\frac{\partial}{\partial m_s} m_N\,.
\label{def-sigmapiN}
\end{eqnarray} 
The sigma terms of the remaining baryon states are defined analogously to  (\ref{def-sigmapiN}). In Tab. \ref{tab:sigmaterms} we 
present our predictions for the pion- and strangeness 
sigma terms of the baryon octet and decuplet states for our parameter sets of Tab. \ref{tab:FitParametersA}.

In Tab. \ref{tab:sigmaterms} the sigma terms for the octet states are compared with two recent lattice determinations \cite{Durr:2011mp,Horsley:2011wr}. 
Our values for the non-strange sigma terms are in reasonable agreement with the lattice results. In particular, we obtain a rather small value for the
pion-nucleon sigma term $\sigma_{\pi N} = 39^{+2}_{-1}$ MeV, which is compatible with the seminal result $\sigma_{\pi N}= (45 \pm 8)$ MeV 
of Gasser, Leutwyler and Sainio in  \cite{Gasser:1990ce}. The size of the pion-nucleon term can be determined from the pion-nucleon
scattering data. It requires a subtle subthreshold extrapolation of the scattering data \cite{Ditsche:2012ja}. Despite the long history 
of the sigma-term physics, the precise determination is still highly controversial (for one of the first reviews see e.g.  \cite{Reya:1974gk}). 
Such a result is also consistent with the recent analysis of the QCDSF collaboration \cite{Bali:2011ks}, which suggests a value $\sigma_{\pi N} =(38 \pm 12)$ MeV.
In contrast there appears to be a slight tension with the recent analysis \cite{Alvarez-Ruso:2013fza} that obtained 
$\sigma_{\pi N} =(52\pm 3 \pm 8)$ MeV based on flavour SU(2) extrapolation of a large selection of lattice data for the nucleon mass.

In Tab. \ref{tab:sigmaterms} we include a first estimate of the uncertainties in the sigma terms from the different parameter 
sets in Tab. I and Tab. III. A full estimate of the systematic uncertainties is beyond the scope of this work. In particular, it 
would require a study of the corrections to the large-$N_c$ sum rules our analysis relies on. For the counter terms required at 
N$^3$LO such results are not available at present.

\begin{table}[t]
\setlength{\tabcolsep}{2.5mm}
\renewcommand{\arraystretch}{0.9}
\begin{center}
\begin{tabular}{l||c|c|c|rr|rr}\hline
 &  \cite{Durr:2011mp} &  \cite{Horsley:2011wr} &  \cite{Ren:2013dzt}             & Tab. 1 &   Error  & Tab. 3  & Error \\ \hline \hline
$\sigma_{\pi N}$ & $39(4) ^{+ 18}_{- 7}  $ &$ 31 (3)(4)$ & 46(2)(12)              &39.3& 0.6& 40.5& 0.5\\
$\sigma_{\pi \Lambda}$ & $29(3)  ^{+11}_{-5} $ &24(3)(4) & 20(2)(13)              &23.1& 0.3& 23.4& 0.5\\
$\sigma_{\pi \Sigma}$ & $ 28(3) ^{+19}_{-3}$ & 21(3)(3)  & 19(2)(6)               &18.3& 0.2& 18.1& 0.5\\
$\sigma_{\pi  \Xi}$ & $16(2)^{+8}_{-3}  $&16(3)(4)       & 6(2)(5)                &5.7& 0.8& 5.7& 1.0\\
\\
$\sigma_{s N}$ & $\;\;34(14)^{+28}_{-24}  $ &$ 71(34)(59)$       & 157(25)(68)    &84.2& 3.3& 100.8& 11.7\\
$\sigma_{s \Lambda}$ & $\;\;90(13)  ^{+24}_{-38} $ &247(34)(69)  & 256(22)(60)    &229.8& 3.3& 234.6& 3.3\\
$\sigma_{s \Sigma}$ & $ 122(15) ^{+25}_{-36}$ & 336(34)(69)      & 270(22)(47)    &355.2& 4.6& 358.9& 6.2\\
$\sigma_{s  \Xi}$ & $156(16)^{+36}_{-38}  $&468(35)(59)          & 369(23)(50)    &368.2& 8.0& 362.9& 13.3\\
\hline
\\
 &  \cite{MartinCamalich:2010fp} & \cite{Semke:2012gs} &  \cite{Ren:2013oaa}      & Tab. 1 &   Error  & Tab. 3 & Error \\ \hline \hline
$\sigma_{\pi \Delta}$   & $55(4)(18) $  & $34(3) $    & $28(1)(8) $               &38.0& 0.6& 39.7& 1.9\\
$\sigma_{\pi \Sigma^*}$ & $39(3)(13) $  & $28(2) $    & $22(2)(9) $               &26.8& 0.6& 27.2& 0.9\\
$\sigma_{\pi \Xi^*}$    & $22(3)(7)  $  & $18(4)  $   & $11(2)(6)  $              &13.7& 0.5& 14.0& 0.9\\
$\sigma_{\pi  \Omega}$  & $\;\;5(2)(1) $& $10(4) $ & $\;\;5(2)(2) $               &4.4& 0.9& 5.0& 0.9 \\
\\
$\sigma_{s \Delta}$   & $\;\;56(24)(1) $ & $\;\;41(41) $ & $\;\;88(22)(3) $       &42.5& 3.8& 63.6& 26.5\\
$\sigma_{s \Sigma^*}$ & $160(28)(7) $    & $211(44) $   & $243(24)(31) $          &193.5& 4.7& 201.8& 14.5\\
$\sigma_{s \Xi^*}$    & $ 274(32)(9)$    & $373(53)$    & $391(24)(67)$           &309.1& 3.9& 308.1&13.5\\
$\sigma_{s  \Omega}$  & $360(34)(26)  $  & $510(50)  $ & $528(26)(101)  $         &431.3& 4.2& 424.9& 14.7 \\ \hline
\end{tabular}
\caption{Pion- and strangeness sigma terms of the baryon octet and decuplet states in units of MeV. A comparison
with various theoretical predictions \cite{Durr:2011mp,Horsley:2011wr,Ren:2013dzt,MartinCamalich:2010fp,Semke:2012gs,Ren:2013oaa} is provided.
We take the average of sigma terms and strangeness contents as implied by the parameter sets of Tab. I and Tab. III separately. The associated errors
identify the maximum deviation from the average within the given sample.  }
\label{tab:sigmaterms}
\end{center}
\end{table}

Our estimate for the strangeness sigma term of the nucleon with $\sigma_{sN} = 84^{+ 28}_{-\;4}$ MeV is significantly larger 
than the lattice average $\sigma_{sN}= 40^{+10}_{-10}$ MeV obtained in \cite{Junnarkar:2013ac}. For the strangeness sigma terms 
of the remaining octet states there appears to be a striking conflict amongst the values obtained by the BMW and QCDSF-UKQCD 
groups. Our values are close to the values of the QCDSF-UKQCD group. The sigma terms for the baryon decuplet states are 
compared with three previous extrapolation results \cite{MartinCamalich:2010fp,Semke:2012gs,Ren:2013oaa}. Our values in 
Tab. \ref{tab:sigmaterms} are consistent with the previous analysis \cite{Semke:2012gs}, based on the same framework used in 
this work. We observe that though the inclusion of finite volume effects is instrumental to determine the values of the symmetry 
preserving counter terms, the overall effect on the baryon decuplet masses at large volumes turns out rather small. There is also 
quite a consistency with the decuplet sigma terms obtained in \cite{MartinCamalich:2010fp,Ren:2013oaa}.

We close with a general comment on the convergence of a chiral expansion of the baryon masses.
So far any strict chiral expansion without a partial summation of higher order terms did
not lead to a satisfactory description of the lattice data. It is an interesting question how small the quark masses 
have to be chosen as to render the heavy-baryon framework meaningful in the flavour $SU(3)$ case. In principal given 
our parameter set we are in a position to identify the power counting regime as studied previously for the two flavour 
case in  \cite{Young:2002ib,Leinweber:2003dg,Leinweber:2005xz}. A detailed analysis of this issue is in preparation 
and will be provided elsewhere.

\clearpage

\section{Summary}
\label{sec:summary}

In this work we reported on a comprehensive analysis of the available three flavour QCD lattice simulations of six different groups on the baryon octet and decuplet masses. We obtained an accurate 12 parameter description of altogether more than  220 lattice data points, where we kept all data with pion masses smaller than about 600 MeV. Our study is based on the relativistic three-flavour chiral Lagrangian with baryon octet and decuplet degrees of freedom.
The baryon self energies were computed in a finite box at N$^3$LO, where the physical masses are kept inside all loop integrals. The low-energy parameters were constrained by using large-$N_c$ sum rules. In contrast to previous works all power-law finite volume corrections are incorporated for the decuplet baryons. We found their effects to be significant.

Predictions for all relevant low-energy parameters were obtained. In partic\-ular we extracted a pion-nucleon sigma term of $39_{-1}^{+2}$ MeV and a strangeness sigma term of the nucleon of
$\sigma_{sN} = 84^{+ 28}_{-\;4}$ MeV. The flavour SU(3) chiral limit of the baryon octet and decuplet masses was determined with $( 802 \pm 4)$ MeV and $(1103 \pm 6)$ MeV. In our fits we used the empirical masses of the baryon octet and decuplet states as a constraint. That allowed us to perform independent scale settings for the various lattice data. We obtained results for the lattice scales that are compatible with previous estimates, but appear to be much more accurate. Detailed predictions for the baryon masses as currently evaluated
by the ETM lattice QCD group are made.

\section*{Acknowledgments}
\begin{acknowledgments}
Financial support from the Thailand Research Fund through the Royal Golden
Jubilee Ph.D. Program (Grant No. PHD/0227/2553) to R. Bavontaweepanya and C. Kobdaj is acknowledged.
C. Kobdaj  was also partially supported by SUT research fund.
M.F.M. Lutz thanks  C. Alexandrou for providing a table with pion and kaon masses of an ongoing ETM simulation
on the baryon masses and G. Schierholz for stimulating discussions on the QCDSF-UKQCD data.
We are grateful to R\"udiger Berlich of Gemfony Scientific UG for help with the optimisation library Geneva.
\end{acknowledgments}
\clearpage

\appendix*
\section*{Appendix A}

We detail the running of the low-energy parameters on the renormalization scale $\mu$, where we
use the following convention

\begin{eqnarray}
\bar g = \hat g - \frac{1}{4}\,\frac{\Gamma_g(\mu) }{(4\,\pi\,f)^2}\, \log \frac{\mu^2}{M^2} \,,
\end{eqnarray}

with all parameters $\hat g$ being scale independent.
We find

\allowdisplaybreaks[1]
\begin{eqnarray}
&& \Gamma_{b_0} = \frac{7\, \Delta \,M (\Delta+2\, M)^3}{72 \,(\Delta+M)^4}\,C^2\,, \qquad \qquad
 \Gamma_{b_D} =-\frac{\Delta \,M \,(\Delta+2 \,M)^3}{24\, (\Delta+M)^4}\,C^2\,,
\nonumber\\
&& \Gamma_{b_F} =\frac{5 \, \Delta\, M\, (\Delta+2\, M)^3}{144 \,(\Delta+M)^4}\,C^2\,,  \qquad \qquad
 \Gamma_{d_0} =-\frac{ \Delta\, (\Delta+2\, M)^3}{48\, M^2\, (\Delta+M)}\,C^2\,,
\nonumber\\
&& \Gamma_{d_D} =-\frac{ \Delta \,(\Delta+2\, M)^3}{48\, M^2\, (\Delta+M)}\,C^2\,,
\nonumber\\ \nonumber\\ \nonumber\\
&& \Gamma_{\zeta_0} =-\frac{8 \left(13 \,D^2+9\, F^2\right) (\Delta+M)^4-7\, C^2 \,(\Delta+2 M)^2
   \left(\Delta^2+4 \,\Delta\, M-2 M^2\right)}{36 \,(\Delta+M)^4} \,,
\nonumber\\
&& \Gamma_{\zeta_D} =\frac{12 \left(D^2-3\, F^2\right) (\Delta+M)^4-C^2 \,(\Delta+2 \,M)^2
   \left(\Delta^2+4 \,\Delta M-2\, M^2\right)}{12\, (\Delta+M)^4}\,,
\nonumber\\
&& \Gamma_{\zeta_F} =-\frac{5 \left(48\, D\, F \,(\Delta+M)^4-C^2\, (\Delta+2 \,M)^2
   \left(\Delta^2+4 \,\Delta M-2\, M^2\right)\right)}{72\,   (\Delta+M)^4}\,,
\nonumber\\
&& \Gamma_{\xi_0} =-\frac{27 \,C^2 \left(4 \,M\, (\Delta+M)^3+3 (\Delta+M)^4+M^4\right)+200 \,H^2\,
   M^2 \,(\Delta+M)^2}{648\, M^2\, (\Delta+M)^2}\,,
\nonumber\\
&& \Gamma_{\xi_D} =-\frac{3 \,C^2 \left(4 \,M \,(\Delta+M)^3+3 \,(\Delta+M)^4+M^4\right)+40 \,H^2\,
   M^2 \,(\Delta+M)^2}{72\, M^2 \,(\Delta+M)^2}\,,
\nonumber\\ \nonumber\\ \nonumber\\
&& \Gamma_{c_0} = \frac{20 \,\bar b_0}{3}+4 \,\bar b_D +
5\,\frac{55\, \gamma^{(0)}_C\,C^2+96 \,(D^2+3\, F^2)}{432 \,M}
\nonumber\\
&& \qquad +\,
\frac{1}{144}\, \Big( -4 \,(30 \,g_0^{(S)}+9 \,g_1^{(S)}+26 \,g_D^{(S)})-\bar M_{[8]}\, (30\,
   g_0^{(V)}+9 \,g_1^{(V)}+26 \,g_D^{(V)})\Big)
\nonumber\\
&& \qquad -\, \frac{2}{27} \left(18\, \hat b_D \left(\gamma^{(4)}_C\,C^2+D^2+3 \,F^2\right)+C^2 \,(15\,
   \gamma^{(4)}_C\,\hat b_F-11 \,\gamma^{(5)}_C \,\hat d_D)+108\, \hat b_F\, D\, F\right)\,,
\nonumber\\
&& \Gamma_{c_1} = \frac{96 \left(D^2-3\, F^2\right)-25 \,\gamma^{(0)}_C\,C^2}{216\, M}+
\frac{1}{24} \,\Big(-4 \,g_1^{(S)}-g_1^{(V)}\, \bar M_{[8]} \Big)
\nonumber\\
&& \qquad -\,\frac{4}{27} \left(6 \,\hat b_D \left(\gamma^{(4)}_C\,C^2+8 \,D^2\right)+5\, C^2\, (3\,
  \gamma^{(4)}_C\, \hat b_F-\gamma^{(5)}_C \,\hat d_D)\right)\,,
\nonumber\\
&& \Gamma_{c_2} = \frac{2 \,\bar b_D}{3}+ \frac{25 \,\gamma^{(0)}_C\, C^2+12\, D^2-36\, F^2}{36\, M}
 + \frac{1}{16} \,\Big(4\, (g_1^{(S)}+g_D^{(S)})+\bar M_{[8]}\, (g_1^{(V)}+g_D^{(V)})\Big)
\nonumber\\
&& \qquad +\,\frac{2}{9} \left(3\, \hat b_D \left(2 \,\gamma^{(4)}_C\,C^2+5 \,D^2+9 \,F^2\right)+C^2 \,(15\,
  \gamma^{(4)}_C\, \hat b_F-8 \,\gamma^{(5)}_C \,\hat d_D)+54\, \hat b_F\, D\, F\right)\,,
\nonumber\\
&& \Gamma_{c_3} = \frac{2 \,\bar b_F}{3}  -\frac{25\, \gamma^{(0)}_C\,C^2+96\, D \,F}{48\, M}
+\frac{1}{16} \,\Big(4\, g_F^{(S)}+g_F^{(V)}\, \bar M_{[8]}\Big )
\nonumber\\
&& \qquad +\,\frac{2}{9} \left(5 \,\hat b_D \left(\gamma^{(4)}_C\,C^2+6 \,D \,F\right)+3\, \hat b_F \left(-2\,
   \gamma^{(4)}_C\,C^2+5\, D^2+9 \,F^2\right)+2\, C^2\,\gamma^{(5)}_C \, \hat d_D\right)\,,
\nonumber\\
&& \Gamma_{c_4} = \frac{44 \,\bar b_D}{9} -2\,\frac{25\, \gamma^{(0)}_C\,C^2+21 \left(D^2-3\, F^2\right)}{27 \,M}
-\frac{1}{72} \,\Big( 36\, g_1^{(S)}+52 \,g_D^{(S)} +  \bar M_{[8]}\,( 9\, g_1^{(V)} + 13\, g_D^{(V)})
  \Big)
\nonumber\\
&& \qquad +\,\frac{4}{27} \left(-9\, \hat b_0\, \gamma^{(4)}_C\,C^2+3 \,\hat b_D \left(\gamma^{(4)}_C\,C^2+8 \,D^2\right)+C^2
   (-15 \,\gamma^{(4)}_C\,\hat b_F+9\,\gamma^{(5)}_C \, \hat d_0+8\,\gamma^{(5)}_C \, \hat d_D)\right)\,,
\nonumber\\
&& \Gamma_{c_5} = \frac{44 \,\bar b_F}{9} +13\,\frac{25 \,\gamma^{(0)}_C\,C^2+96\, D\, F }{216 \,M}
-\frac{13}{72} \,\Big(4 g_F^{(S)}+g_F^{(V)}\, \bar M_{[8]}\Big)
\nonumber\\
&& \qquad +\,\frac{2}{27}\, C^2\, (15\, \gamma^{(4)}_C\,\hat b_0+42\, \gamma^{(4)}_C\,\hat b_F
-15\,\gamma^{(5)}_C \, \hat d_0-19\, \gamma^{(5)}_C \,\hat d_D)\,,
\nonumber\\
&& \Gamma_{c_6} = \frac{44 \,\bar b_0}{9}+ \frac{875 \,\gamma^{(0)}_C\,C^2+896\, D^2-576\, F^2}{432 \,M}
\nonumber\\
&& \qquad +\,\frac{1}{432}\,\Big(-264 \,g_0^{(S)}+108\, g_1^{(S)} +32\, g_D^{(S)} +  \bar M_{[8]}\,(-66 \,g_0^{(V)}+27\, g_1^{(V)}\,
  +8\, g_D^{(V)} )\Big)
\nonumber\\
&& \qquad +\,\frac{2}{27} \left(42\, \hat b_0 \,\gamma^{(4)}_C\,C^2+18 \,\hat b_D \left(\gamma^{(4)}_C\,C^2-2\, D^2\right)+C^2
   (15 \,\gamma^{(4)}_C\,\hat b_F-42\, \gamma^{(5)}_C \,\hat d_0-19 \,\gamma^{(5)}_C \,\hat d_D)\right)\,,
\nonumber\\ \nonumber\\ \nonumber\\
&& \Gamma_{e_0} = \frac{20 \,\bar d_0}{3}+2 \,\bar d_D+\frac{81 \,\gamma^{(1)}_C\,C^2+196 \,\gamma^{(0)}_H\,H^2}{432\, M}
\nonumber\\
&& \qquad +\, \frac{1}{72}\,\Big(-60\, \tilde h_1^{(S)}-52 \,\tilde h_2^{(S)}-36\, \tilde h_3^{(S) }
-\bar M_{[10]} \,(15 \,h_1^{(V)}+13\, h_2^{(V)} +9\,   h_3^{(V)})\Big)
\nonumber\\
&& \qquad+\,  \frac{2}{3} \,C^2 \,\gamma^{(2)}_C\,(\hat b_D-\hat b_F)-\frac{10\, \hat d_D H^2}{81}\,,
\nonumber\\
&& \Gamma_{e_1} = \frac{27 \,\gamma^{(1)}_C\,C^2-56 \,\gamma^{(0)}_H\,H^2}{216\, M} +
\frac{1}{12} \,\Big(-4 \,\tilde h_3^{(S)}-h_3^{(V)} \bar M_{[10]}\Big)
+\frac{4}{9} \,C^2 \,(3\,\gamma^{(2)}_C\, \hat b_F-\gamma^{(3)}_C\,\hat d_D) \,,
\nonumber\\
&& \Gamma_{e_2} = \frac{2 \,\bar d_D}{3}-\frac{27\, \gamma^{(1)}_C\,C^2+28\, \gamma^{(0)}_H\,H^2}{144\, M}
+ \frac{1}{8}\,\Big(4\, (\tilde h_2^{(S)}+\tilde h_3^{(S)})+\bar M_{[10]}\,
   (h_2^{(V)}+h_3^{(V)})\Big)
 \nonumber\\
&& \qquad+\,\frac{2}{27} \left(-18\,C^2 \gamma^{(2)}_C\,\hat b_D \,+9 \,C^2 \,\gamma^{(3)}_C\,\hat d_D+5\, \hat d_D\,
   H^2\right)\,,
\nonumber\\
&& \Gamma_{e_3} = \frac{44 \,\bar d_D}{9}+\frac{81 \,\gamma^{(1)}_C\,C^2+196\, \gamma^{(0)}_H\,H^2}{216\, M}-
\frac{1}{36} \,\Big( 52 \,\tilde h_2^{(S)} 36\, \tilde h_3^{(S)}+\bar M_{[10]}\,( 13\, h_2^{(V)} +9\,
   h_3^{(V)} )\Big)
 \nonumber\\
&& \qquad-\, \frac{2}{3} \,C^2 (\gamma^{(2)}_C\,\hat b_0+2 \,\gamma^{(2)}_C\,\hat b_F-\gamma^{(3)}_C\,\hat d_0-\gamma^{(3)}_C\,\hat d_D)\,,
\nonumber\\
&& \Gamma_{e_4} = \frac{44 \,\bar d_0}{9}-\frac{7 \,\gamma^{(0)}_H\,H^2}{81\, M}+
\frac{1}{216}\,\Big(-132 \,\tilde h_1^{(S)}+16 \,\tilde h_2^{(S)} -33 \,h_1^{(V)} \bar M_{[10]}+4\,
   h_2^{(V)} \bar M_{[10]}\Big)
  \nonumber\\
&& \qquad-\,   \frac{2}{3} \,C^2\, (\gamma^{(2)}_C\,\,\hat b_0+\gamma^{(2)}_C\,\,\hat b_D-\gamma^{(2)}_C\,\,\hat b_F
-\gamma^{(3)}_C\,\hat d_0)\,,
\end{eqnarray}
where all $\gamma$'s approach $1$ in the limit $\Delta \to 0$. It holds
\begin{eqnarray}
&& \gamma^{(0)}_C =\frac{8\,\Delta^5 + 48\,\Delta^4\,M + 120\,\Delta^3\,M^2 +
      156\,\Delta^2\,M^3 + 102\,\Delta\,M^4 + 25\,M^5}{ 25 \,(M+ \Delta)^6}\,M\,,
\nonumber\\
&& \gamma^{(1)}_C  = \frac{(3\,M+ \Delta)^2}{9\,M^3}\,(M- \Delta)\,,\qquad  \quad
\gamma^{(2)}_C = \,\frac{(\Delta + 2 M)^2 \,(\Delta^2 + \Delta \,M + M^2)}{4\,M^3\,(\Delta +M) }\,,
\qquad \qquad
\nonumber\\
&& \gamma^{(3)}_C = \,\frac{(\Delta + 2 M)^2 \,(3\,\Delta^2 + 4\,\Delta \,M + 2\,M^2)}{8\,M^2\,(\Delta +M)^2 }\,,\quad  \qquad
\gamma^{(5)}_C = \frac{(-\Delta + M)\,M^2\, (\Delta + 2\, M)^2}{4\,(\Delta +M)^5 }\,,
\nonumber\\
&&\gamma^{(4)}_C = \frac{(-\Delta^2 - 4 \,\Delta\, M + 2\, M^2)\, (\Delta + 2\, M)^2}{8\,(\Delta +M)^4 } \,, \qquad \quad
 \gamma^{(0)}_H = \frac{M}{M+ \Delta}\,,
\nonumber\\ \nonumber\\
&&\tilde h_1^{(S)} = h_1^{(S)} + h_2^{(S)}/4\,, \qquad
\tilde h_2^{(S)} = h_3^{(S)} + h_4^{(S)}/4\,, \qquad
\tilde h_3^{(S)} = h_5^{(S)} + h_6^{(S)}/4\,.
\end{eqnarray}
Note that the running of the symmetry breaking counter terms $\Gamma_{c_i}$ and $\Gamma_{e_i}$ includes the effect
of the wave function factors $\Gamma_\zeta$ and $\Gamma_\xi$.
\bibliography{1}
\bibliographystyle{apsrev4-1}
\end{document}